\documentclass[%
 reprint,
superscriptaddress,
 amsmath,amssymb,
 aps,
prd,
floatfix
]{revtex4-2}
\usepackage{graphicx}
\usepackage{dcolumn}
\usepackage{bm}
\usepackage{footnote}

\usepackage[english]{babel}
\usepackage{xspace}
\usepackage[per-mode=symbol]{siunitx}
\usepackage[utf8]{inputenc}
\usepackage{multirow}
\usepackage{placeins}

\newcommand{\Xmax}{$\textit{X}_\mathrm{max}$~}
\newcommand{\logE}{$\text{log}_{10}(\textit{S}_{125}\text{/VEM})$}
\newcommand{\Deg}{$^\circ$}
\newcommand{\classifier}{\emph{Classifier}}
\newcommand{\denoiser}{\emph{Denoiser}}

\begin{document}


\title{Identification and Denoising of Radio Signals from Cosmic-Ray\\ Air Showers using Convolutional Neural Networks}

\affiliation{III. Physikalisches Institut, RWTH Aachen University, D-52056 Aachen, Germany}
\affiliation{Department of Physics, University of Adelaide, Adelaide, 5005, Australia}
\affiliation{Dept. of Physics and Astronomy, University of Alaska Anchorage, 3211 Providence Dr., Anchorage, AK 99508, USA}
\affiliation{School of Physics and Center for Relativistic Astrophysics, Georgia Institute of Technology, Atlanta, GA 30332, USA}
\affiliation{Dept. of Physics, Southern University, Baton Rouge, LA 70813, USA}
\affiliation{Dept. of Physics, University of California, Berkeley, CA 94720, USA}
\affiliation{Lawrence Berkeley National Laboratory, Berkeley, CA 94720, USA}
\affiliation{Institut f{\"u}r Physik, Humboldt-Universit{\"a}t zu Berlin, D-12489 Berlin, Germany}
\affiliation{Fakult{\"a}t f{\"u}r Physik {\&} Astronomie, Ruhr-Universit{\"a}t Bochum, D-44780 Bochum, Germany}
\affiliation{Universit{\'e} Libre de Bruxelles, Science Faculty CP230, B-1050 Brussels, Belgium}
\affiliation{Vrije Universiteit Brussel (VUB), Dienst ELEM, B-1050 Brussels, Belgium}
\affiliation{Dept. of Physics, Simon Fraser University, Burnaby, BC V5A 1S6, Canada}
\affiliation{Department of Physics and Laboratory for Particle Physics and Cosmology, Harvard University, Cambridge, MA 02138, USA}
\affiliation{Dept. of Physics, Massachusetts Institute of Technology, Cambridge, MA 02139, USA}
\affiliation{Dept. of Physics and The International Center for Hadron Astrophysics, Chiba University, Chiba 263-8522, Japan}
\affiliation{Department of Physics, Loyola University Chicago, Chicago, IL 60660, USA}
\affiliation{Dept. of Physics and Astronomy, University of Canterbury, Private Bag 4800, Christchurch, New Zealand}
\affiliation{Dept. of Physics, University of Maryland, College Park, MD 20742, USA}
\affiliation{Dept. of Astronomy, Ohio State University, Columbus, OH 43210, USA}
\affiliation{Dept. of Physics and Center for Cosmology and Astro-Particle Physics, Ohio State University, Columbus, OH 43210, USA}
\affiliation{Niels Bohr Institute, University of Copenhagen, DK-2100 Copenhagen, Denmark}
\affiliation{Dept. of Physics, TU Dortmund University, D-44221 Dortmund, Germany}
\affiliation{Dept. of Physics and Astronomy, Michigan State University, East Lansing, MI 48824, USA}
\affiliation{Dept. of Physics, University of Alberta, Edmonton, Alberta, T6G 2E1, Canada}
\affiliation{Erlangen Centre for Astroparticle Physics, Friedrich-Alexander-Universit{\"a}t Erlangen-N{\"u}rnberg, D-91058 Erlangen, Germany}
\affiliation{Physik-department, Technische Universit{\"a}t M{\"u}nchen, D-85748 Garching, Germany}
\affiliation{D{\'e}partement de physique nucl{\'e}aire et corpusculaire, Universit{\'e} de Gen{\`e}ve, CH-1211 Gen{\`e}ve, Switzerland}
\affiliation{Dept. of Physics and Astronomy, University of Gent, B-9000 Gent, Belgium}
\affiliation{Dept. of Physics and Astronomy, University of California, Irvine, CA 92697, USA}
\affiliation{Karlsruhe Institute of Technology, Institute for Astroparticle Physics, D-76021 Karlsruhe, Germany}
\affiliation{Karlsruhe Institute of Technology, Institute of Experimental Particle Physics, D-76021 Karlsruhe, Germany}
\affiliation{Dept. of Physics, Engineering Physics, and Astronomy, Queen's University, Kingston, ON K7L 3N6, Canada}
\affiliation{Department of Physics {\&} Astronomy, University of Nevada, Las Vegas, NV 89154, USA}
\affiliation{Nevada Center for Astrophysics, University of Nevada, Las Vegas, NV 89154, USA}
\affiliation{Dept. of Physics and Astronomy, University of Kansas, Lawrence, KS 66045, USA}
\affiliation{Centre for Cosmology, Particle Physics and Phenomenology - CP3, Universit{\'e} catholique de Louvain, Louvain-la-Neuve, Belgium}
\affiliation{Department of Physics, Mercer University, Macon, GA 31207-0001, USA}
\affiliation{Dept. of Astronomy, University of Wisconsin{\textemdash}Madison, Madison, WI 53706, USA}
\affiliation{Dept. of Physics and Wisconsin IceCube Particle Astrophysics Center, University of Wisconsin{\textemdash}Madison, Madison, WI 53706, USA}
\affiliation{Institute of Physics, University of Mainz, Staudinger Weg 7, D-55099 Mainz, Germany}
\affiliation{Department of Physics, Marquette University, Milwaukee, WI 53201, USA}
\affiliation{Institut f{\"u}r Kernphysik, Universit{\"a}t M{\"u}nster, D-48149 M{\"u}nster, Germany}
\affiliation{Bartol Research Institute and Dept. of Physics and Astronomy, University of Delaware, Newark, DE 19716, USA}
\affiliation{Dept. of Physics, Yale University, New Haven, CT 06520, USA}
\affiliation{Columbia Astrophysics and Nevis Laboratories, Columbia University, New York, NY 10027, USA}
\affiliation{Dept. of Physics, University of Oxford, Parks Road, Oxford OX1 3PU, United Kingdom}
\affiliation{Dipartimento di Fisica e Astronomia Galileo Galilei, Universit{\`a} Degli Studi di Padova, I-35122 Padova PD, Italy}
\affiliation{Dept. of Physics, Drexel University, 3141 Chestnut Street, Philadelphia, PA 19104, USA}
\affiliation{Physics Department, South Dakota School of Mines and Technology, Rapid City, SD 57701, USA}
\affiliation{Dept. of Physics, University of Wisconsin, River Falls, WI 54022, USA}
\affiliation{Dept. of Physics and Astronomy, University of Rochester, Rochester, NY 14627, USA}
\affiliation{Department of Physics and Astronomy, University of Utah, Salt Lake City, UT 84112, USA}
\affiliation{Dept. of Physics, Chung-Ang University, Seoul 06974, Republic of Korea}
\affiliation{Oskar Klein Centre and Dept. of Physics, Stockholm University, SE-10691 Stockholm, Sweden}
\affiliation{Dept. of Physics and Astronomy, Stony Brook University, Stony Brook, NY 11794-3800, USA}
\affiliation{Dept. of Physics, Sungkyunkwan University, Suwon 16419, Republic of Korea}
\affiliation{Institute of Physics, Academia Sinica, Taipei, 11529, Taiwan}
\affiliation{Dept. of Physics and Astronomy, University of Alabama, Tuscaloosa, AL 35487, USA}
\affiliation{Dept. of Astronomy and Astrophysics, Pennsylvania State University, University Park, PA 16802, USA}
\affiliation{Dept. of Physics, Pennsylvania State University, University Park, PA 16802, USA}
\affiliation{Dept. of Physics and Astronomy, Uppsala University, Box 516, SE-75120 Uppsala, Sweden}
\affiliation{Dept. of Physics, University of Wuppertal, D-42119 Wuppertal, Germany}
\affiliation{Deutsches Elektronen-Synchrotron DESY, Platanenallee 6, D-15738 Zeuthen, Germany}

\author{R. Abbasi}
\affiliation{Department of Physics, Loyola University Chicago, Chicago, IL 60660, USA}
\author{M. Ackermann}
\affiliation{Deutsches Elektronen-Synchrotron DESY, Platanenallee 6, D-15738 Zeuthen, Germany}
\author{J. Adams}
\affiliation{Dept. of Physics and Astronomy, University of Canterbury, Private Bag 4800, Christchurch, New Zealand}
\author{S. K. Agarwalla}
\thanks{also at Institute of Physics, Sachivalaya Marg, Sainik School Post, Bhubaneswar 751005, India}
\affiliation{Dept. of Physics and Wisconsin IceCube Particle Astrophysics Center, University of Wisconsin{\textemdash}Madison, Madison, WI 53706, USA}
\author{J. A. Aguilar}
\affiliation{Universit{\'e} Libre de Bruxelles, Science Faculty CP230, B-1050 Brussels, Belgium}
\author{M. Ahlers}
\affiliation{Niels Bohr Institute, University of Copenhagen, DK-2100 Copenhagen, Denmark}
\author{J.M. Alameddine}
\affiliation{Dept. of Physics, TU Dortmund University, D-44221 Dortmund, Germany}
\author{S. Ali}
\affiliation{Dept. of Physics and Astronomy, University of Kansas, Lawrence, KS 66045, USA}
\author{N. M. Amin}
\affiliation{Bartol Research Institute and Dept. of Physics and Astronomy, University of Delaware, Newark, DE 19716, USA}
\author{K. Andeen}
\affiliation{Department of Physics, Marquette University, Milwaukee, WI 53201, USA}
% ...
\date{\today}
\date{\today}

\collaboration{IceCube Collaboration}
\noaffiliation

\date{\today}

\begin{abstract}
Radio pulses generated by cosmic-ray air showers can be used to reconstruct key properties like the energy and depth of the electromagnetic component of cosmic-ray air showers. Radio detection threshold, influenced by natural and anthropogenic radio background, can be reduced through various techniques. 
In this work, we demonstrate that convolutional neural networks (CNNs) are an effective way to lower the threshold. We developed two CNNs: a classifier to distinguish radio signal waveforms from background noise and a denoiser to clean contaminated radio signals. Following the training and testing phases, we applied the networks to air-shower data triggered by scintillation detectors of the prototype station for the enhancement of IceTop, IceCube's surface array at the South Pole. Over a four-month period, we identified 554 cosmic-ray events in coincidence with IceTop, approximately five times more compared to a reference method based on a cut on the signal-to-noise ratio. Comparisons with IceTop measurements of the same air showers confirmed that the CNNs reliably identified cosmic-ray radio pulses and outperformed the reference method. 
Additionally, we find that CNNs reduce the false-positive rate of air-shower candidates and effectively denoise radio waveforms, thereby improving the accuracy of the power and arrival time reconstruction of radio pulses.

\end{abstract}

\maketitle

\section{Introduction}\label{sec:intro}

High-energy cosmic rays interacting with Earth's atmosphere generate extensive air showers. These showers comprise numerous charged particles that emit a coherent radio pulse, primarily due to the deflection of electrons and positrons by the Earth's magnetic field~\cite{Huege:2016veh,Schroder:2016hrv}. Detecting this impulsive radio emission has proven valuable for reconstructing air-shower properties, such as electromagnetic energy and depth of the shower maximum, \Xmax \cite{LOPES:2021ipp,Buitink:2014eqa,PierreAuger:2016vya,Bezyazeekov:2018yjw,PierreAuger:2023lkx}. 

Complementary measurements of air-shower events using radio and other detection techniques, particularly particle detectors sensitive to the muonic component, further enhance sensitivity to the mass of the primary particle~\cite{Holt:2019fnj,Flaggs:2023exc,Andringa:2011zz,KASCADE:2005ynk}.
The radio technique is thus expected to play an important role in next-generation experiments targeting the particle and astrophysics of ultra-high-energy cosmic rays~\cite{Coleman:2022abf}.

A primary challenge of using radio signals for air-shower measurements is continuous background noise from natural (Galactic and thermal) and anthropogenic sources.
The system noise is often small compared to the external background, e.g., for the SKA Log-periodic antennas (SKALA) v2 and their low-noise amplifiers used here, the internal noise is approximately $40\,$K~\cite{7297231,benthem2021aperture}, only.
This noise contaminates radio measurements, obscuring air-shower pulses and limiting the energy threshold for cosmic-ray studies.

Several techniques have been employed by radio experiments to mitigate the impact of background noise on air-shower measurements. 
One approach is digital radio interferometry~\cite{LOPES:2021ipp,ANITA:2010ect,Schoorlemmer:2020low}, which combines measurements from multiple antennas to improve the signal-to-noise ratio (SNR).
Another method involves selecting an optimal frequency band ~\cite{BalagopalV:2017aan}: 
while air-shower radio emissions are broadband (from MHz to GHz), Galactic radio background dominates below a few 100\,MHz. Thus, selecting an optimal frequency range can significantly enhance the SNR.
Moreover, narrow-band radio frequency interferences, which are a major component of anthropogenic background, can be effectively filtered using notch filters and similar techniques, because the air-shower radio signals are broad-band. 
More recently, machine learning tools (see~\cite{hepmllivingreview} for a living review) have been introduced to further reduce background noise. For example, Convolutional Neural Networks (CNNs) have shown promise in denoising contaminated radio traces when trained on simulated~\cite{Erdmann:2019nie} and measured background~\cite{Bezyazeekov:2015rpa,Shipilov:2018wph,Bezyazeekov:2021sha}. 
Building on these advancements, this work presents the development of two independent CNNs: a \classifier\ for identifying traces containing radio pulses and a \denoiser\ for recovering underlying air-shower pulses from noisy traces~\cite{A-Rehman:2024}.

The performance of these CNNs is demonstrated by application to radio measurements of air showers from IceCube's surface array at the South Pole.
The surface component of IceCube, known as IceTop, consists of 81 stations, each with pairs of ice-Cherenkov tanks, spread over an area of approximately 1\,km$^2$. 
IceTop is designed to detect cosmic-ray air showers in the energy range from below a PeV to a few EeV~\cite{IceCube:2012nn}. 
However, snow accumulation on IceTop tanks has increased the detection threshold and introduced systematic uncertainties in air-shower measurements over time~\cite{IceCube:2023nyj}.
To improve IceTop's detection threshold and expand its scientific capabilities, such as precise energy calibration and improved composition studies, a surface array enhancement of IceTop has been planned~\cite{Leszczynska:2019ahq, Haungs:2019ylq, Schroder:2018dvb}. 
This enhancement involves deploying elevated scintillation panels and radio antennas across the existing IceTop array.

While a low false-positive rate is critical for self-triggered radio experiments, externally triggered radio setups, such as the IceTop enhancement, focus primarily on improving the measurement quality of air showers.
Every triggered event is an air shower, but in many the radio signal is too weak and buried in the noise. 
The goal is thus to measure that radio signal in as many air-shower events as possible and to reduce the measurement uncertainty on important parameters of the radio pulse in each antenna, in particular, its arrival time and strength.

\begin{figure}[h]
\centering
\includegraphics[width=0.45\textwidth,angle=-90]{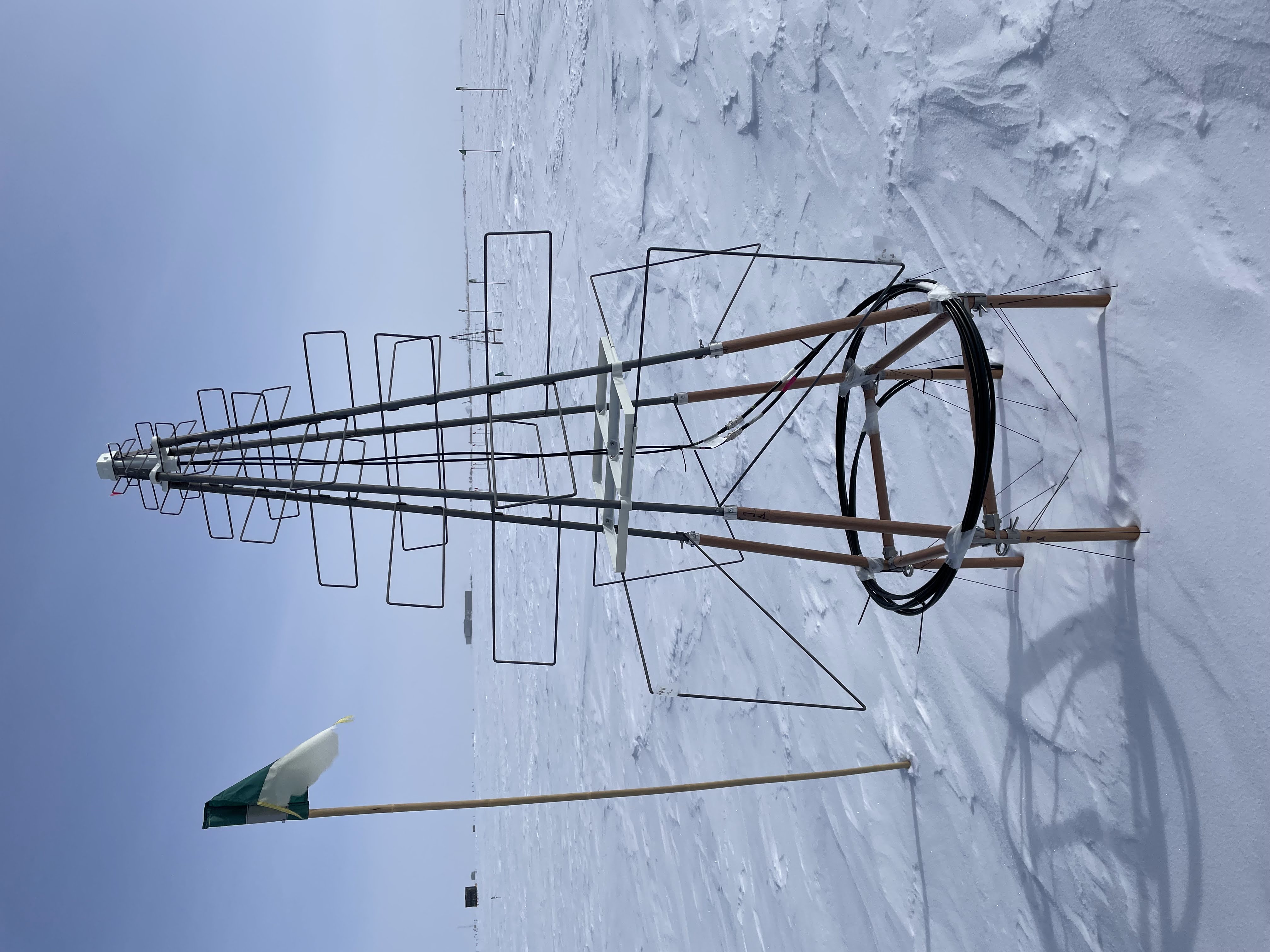}
\caption{One of the three SKALA antennas of the prototype surface station at the South Pole (photo from December 2022, courtesy of Roxanne Turcotte \cite{Turcotte-Tardif:2022smf}).}
\label{fig:SKALA}
\end{figure}

A prototype station of the IceTop enhancement consisting of eight panels and three SKALA v2 antennas was deployed in 2020 and has regularly been measuring air showers with both the scintillation detectors and the radio antennas \cite{IceCube:2021epf} (cf.~Fig.~\ref{fig:SKALA}). 
Since its deployment, the prototype station's firmware and hardware have undergone several upgrades, including an update to the local data acquisition in January 2022~\cite{Turcotte-Tardif:2022smf, IceCube:2023pjc}. 
The radio measurements are broadband ($70-350\,$MHz) with two polarization channels recorded per antenna.
We collect two types of data: air-shower data, triggered by scintillation panels, and fixed-rate-triggered (FRT) data, collected by a software trigger at regular intervals and providing a measurement of the radio background.
For this study, FRT radio data are used to train and test the CNNs, which are then applied to air-shower data to experimentally demonstrate their performance.

This paper details the architecture, training, testing, and implementation of the networks, and is organized as follows: 
Sec.~\ref{sec:architecture} provides the basic architecture of the CNNs. 
Sec.~\ref{sec:dataPrep} covers data preparation for network training and validation using simulated radio pulses and measured background, followed by Sec.~\ref{sec:network_results} on the performance of the networks evaluated on an independent test dataset. 
Sec.~\ref{sec:application} describes applying the networks to measured radio data from the prototype station to identify air-shower events. Sec.~\ref{sec:comparison}, compares CNN results with a standard SNR method for identifying air-shower events. Finally, Sec.~\ref{sec:outlook} provides a summary and outlook.

\section{Network Architecture}\label{sec:architecture}

\begin{figure}[ht!]
\centering
\includegraphics[width=0.48\textwidth]{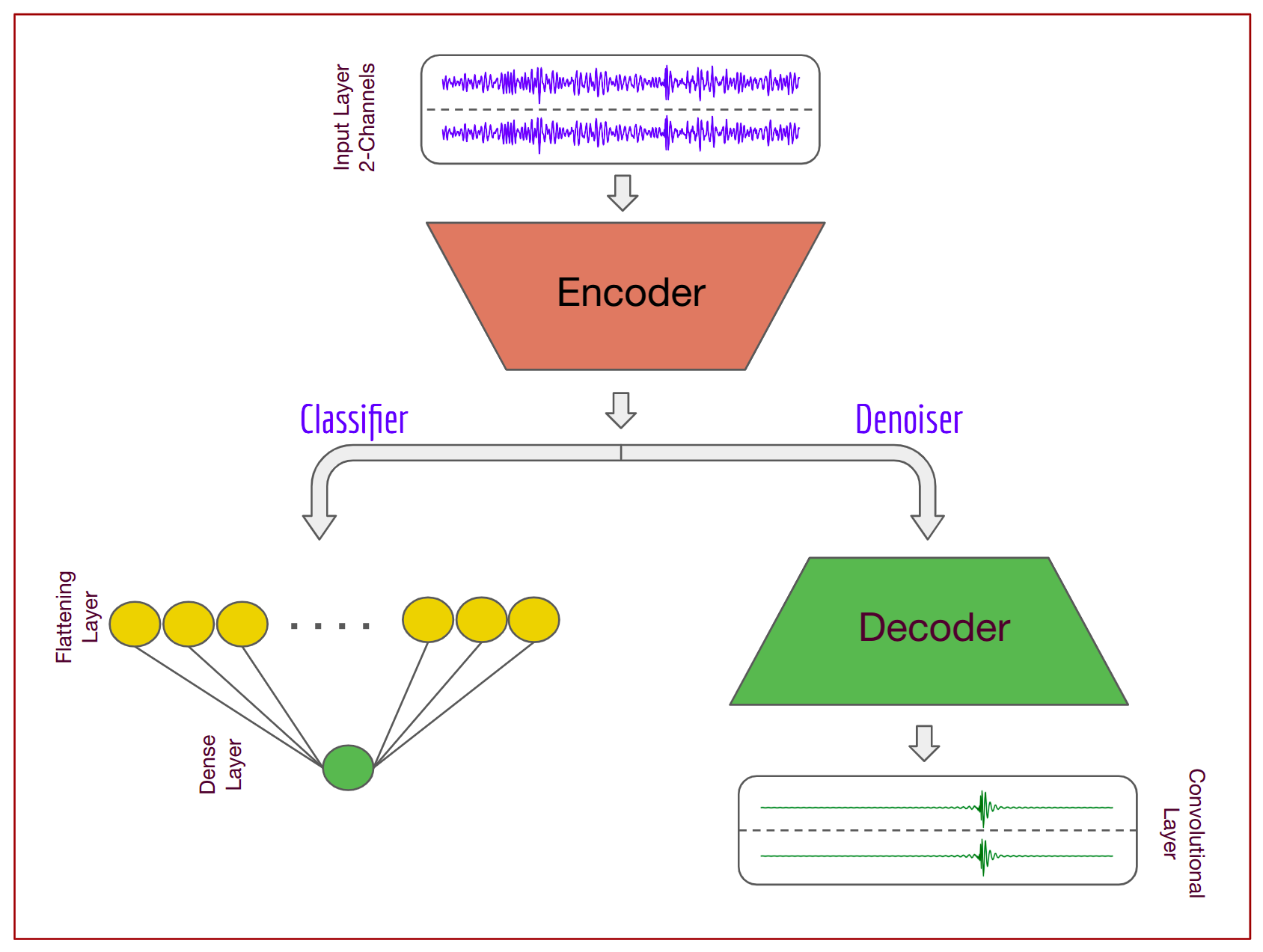}
\caption{Schematic of the neural network architectures for the \classifier\ and \denoiser. Both networks take two-channel input waveforms and begin with an encoder composed of 1D convolutional layers paired with max-pooling layers. In the \classifier, the encoder is followed by a flattening layer and a dense output layer. In contrast, the \denoiser\ includes a decoder composed of up-sampling and convolutional layers, followed by a final convolutional output layer.}
\label{fig:architecture}
\end{figure}

We developed two networks: a \classifier, to distinguish signal traces from noise-only traces, and a \denoiser, to remove noise from contaminated traces.
Both networks take the two-polarization-channel waveforms from an antenna as input and are primarily built from 1D convolutional layers. Their basic architecture is depicted in Fig.~\ref{fig:architecture}. 

Each network's input is a data array of shape $1000\times2$, representing \(1000\) time samples per trace across \(2\) polarization channels.
For both networks, the \emph{input} layer is followed by an encoder composed of several encoding blocks, each consisting of a 1D convolutional layer and a max-pooling layer with a fixed pooling size of $2$ \cite{2014arXiv1412.6071G}.

For the \classifier, extensive hyperparameter testing led to a final architecture with four encoding layers. Each convolutional layer uses 8 filters with a kernel size of 33 bins (33\,ns), followed by a flattening layer and a dense output layer with a single neuron. ReLU activation functions are applied in all layers except the output layer, which uses a sigmoid activation to map the output to a value between 0 (background-like) and 1 (signal-like). This configuration results in a total of 7,393 trainable parameters.

The \denoiser\ adopts an autoencoder architecture, where the input is first compressed by the encoder (of the same architecture as for the classier, but trained separately) and then reconstructed to its original dimensions by the decoder. 
The decoder mirrors the encoder in structure, using the same number of decoding blocks to ensure output dimensionality matches the input. Each decoding block includes a 1D convolutional layer followed by an upsampling layer with a factor of 2, doubling the output size at each step. Each convolutional layer contains 64 filters with a kernel size of 33. A final convolutional layer with one filter serves as the output of the Denoiser. ReLU activation is applied to all layers except the final one, which uses a linear activation. The output retains the same shape as the input radio traces, with 1000 samples per polarization channel, with noise ideally removed from both channels. This architecture has a total of 684,674 trainable parameters. 

The networks were implemented and trained using Keras~\cite{franccois2019keras} and Tensorflow~\cite{abadi2016tensorflow}. 
Binary cross-entropy (BCE) served as the loss function for the \classifier, while mean squared error (MSE) was used for the \denoiser.
Training was monitored using an early stopping criterion based on the validation loss, with a patience of 10 epochs. If no improvement in validation loss occurred for 10 consecutive epochs, training was halted, and the model weights corresponding to the lowest validation loss were retained (The network architecture and associated processing scripts are available at: \url{https://github.com/icecube/RadioPulseDenoising}).

\section{Data Preparation}\label{sec:dataPrep}

The neural networks were trained using simulated radio signals and measured background data. 
This choice was made because the CoREAS simulations have been shown to accurately describe measured radio pulses from air showers, meaning that no significant discrepancies with experimental data are known~\cite{LOPES:2015eya,T-510:2015pyu,Tunka-Rex:2016nto} . 
However, modeled background often oversimplifies the reality of radio experiments, which is why we used radio background recorded with the exact setup at the South Pole.

CORSIKA \cite{heck1998corsika} v7.7401, with the CoREAS extension ~\cite{huege2013simulating}, was used to generate air-shower simulations and their radio signals. The simulations utilized the average April South Pole atmosphere, which closely approximates the yearly average, with the Fluka model \cite{BOHLEN2014211} for low-energy hadronic interactions and Sibyll 2.3d \cite{Riehn:2019jet} for high-energy interactions. Protons and iron nuclei were used as primary particles, with energies ranging from $10^{16.5}$ to $10^{18.0}$\,eV, zenith angles ($\theta$) from $0$ to $0.8$ in $\sin^2 \theta$ ($0^\circ$ to $63.4^\circ$), and uniform azimuth angles ($\phi$). More than 40,000 distinct CoREAS simulations were generated, wherein radio emissions were computed on a spatial grid configured as an eight-legged star. The center of this star was positioned at the intersection of the shower axis with the ground \cite{IceCube:2022dcd}.

For the radio background, we used FRT event waveforms recorded during the first half of 2022, a period of stable configuration of the updated local data-acquisition.
Each FRT event contains waveforms of $1024$ data points per polarization channel of each of the three antennas of the prototype station, of which the inner $1000$ data points (corresponding to $1\,$\textmu s of radio background) were used in this study.
The total data collection period analyzed in this work amounts to approximately 106 days of effective runtime, during which over 300,000 background events were recorded and used for data preparation.

To prepare the training data, CoREAS radio signals are combined with FRT waveforms and processed using custom software developed within the IceCube framework for radio air-shower analysis \cite{IceCube:2022dcd}. The CoREAS radio pulses are transformed into mock data which include all known properties of the experimental setup. 
As a first step, the CoREAS pulses are zero-padded in the time domain to make all waveforms equal in length. 
These zero-padded signals are then convolved with the SKALA v2 antenna response. 
Next, the convolved signals are resampled from the native $0.2\,$ns to the $1\,$ns sampling rate used for data taking in the prototype station. 
The resampled signals are further convolved with the full electronics response, based on calibration measurements, including components such as a Low-Noise Amplifier and an analog radio board. They are then digitized at 14 bits with a 1\,V peak-to-peak dynamic range (discrete amplitude bins of $2^{-14}$\,V). 
Measured background waveforms in the same units are added to the digitized signals, and the combined traces are converted back to voltages.
Next, pedestal offsets, corresponding to unphysical DC components, are removed from the waveforms, and a box frequency filter is applied to the nominal measurement band of the prototype station of $70-350\,$MHz~\cite{benthem2021aperture}. 
Finally, the electronics response is deconvolved, and the resulting waveforms are saved as the noisy dataset. 

The pure simulated signals (before noise addition) are also filtered to $70-350\,$MHz and used as target labels for the \denoiser. Background waveforms, processed in the same way, are saved for training the \classifier. Each waveform is individually normalized to the range [-1, 1] prior to being passed to the networks to ensure consistent scaling of input features across network layers."
To prevent any positional bias, each simulated radio pulse is randomly time-shifted within the $1\,$\textmu s (1000-bin) waveform before being added to the background.

Following this procedure, approximately 165,000 samples of each type (noisy, pure signal, and background noise) were generated per antenna. Each sample consists of two time series, corresponding to the two antenna polarization channels. Thus, for training and validation of the \denoiser, approximately 165,000 noisy samples paired with their corresponding pure signal were used. For the \classifier, about 165,000 noisy samples and an equal number of background samples were utilized. In total, for each antenna, the complete dataset consists of approximately 495,000 samples.

This processed dataset was split into two subsets: 80\% for training and 20\% for validation. 
After finalizing the networks, an independent dataset was generated following the same procedure to test the network's performance. This included approximately 3,000 new CORSIKA simulations, combined with measured background from the same period. The resulting test set contains approximately 24,000 samples of each type. 

A key quantity to characterize a noisy radio trace is the SNR.
Since definitions vary across experiments, we adopt a version previously used in other experiments, but made it more robust against occasional outliers in the traces by using the median noise level over several time intervals instead of a single time interval~\cite{Rehman:2023jme, Schroder:2023sam}.

\begin{figure}[hbt!]
\centering
\includegraphics[width=0.42\textwidth]{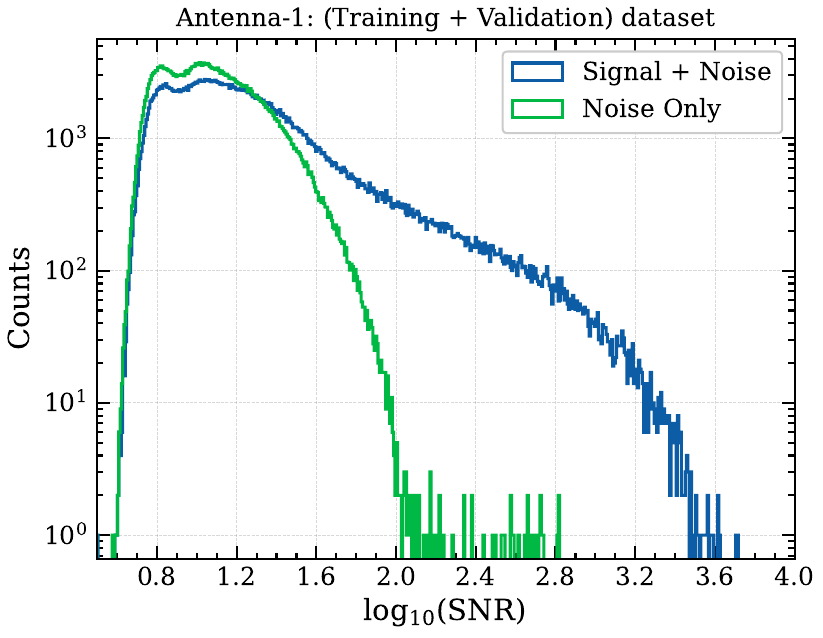}
\includegraphics[width=0.42\textwidth]{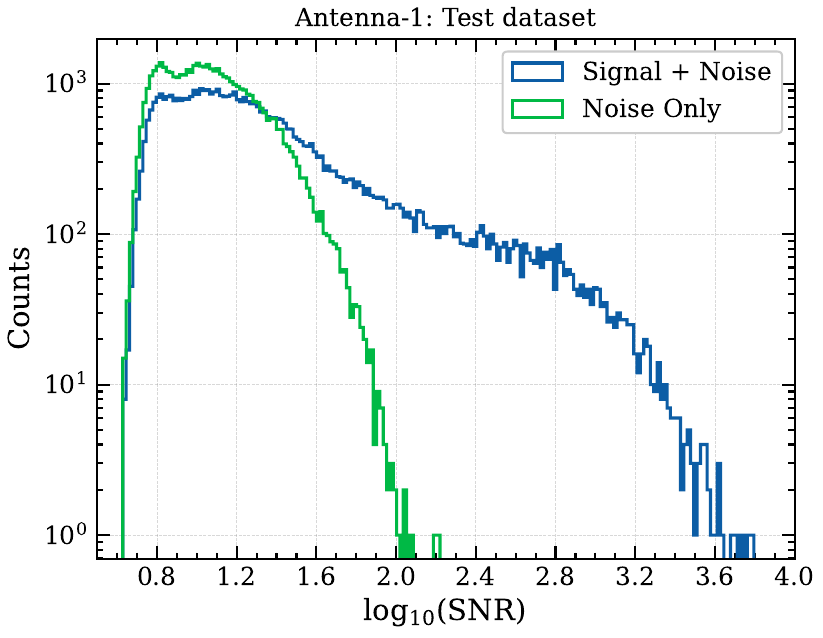}
\caption{Distribution of signal-to-noise ratio of the data set used for training and validation (top) and test (bottom). Shown here is only the distribution for the dataset produced for antenna 1, distributions for other antennas exhibit similar structures, differing only in their mean and slight variations in the tails. The mean background SNR is approximately 14 for both datasets.}
\label{fig:snr_dist}
\end{figure}

The SNR of a voltage trace with amplitudes $Y_i$ in each bin $i$ is calculated as follows:
\begin{equation}\label{eq:snr}
    \text{SNR} = \left( \frac{{\rm max}[|Y_i|]}{{\rm median}[{\rm RMS} (n_j)]} \right)^2~,
\end{equation}
where the numerator is the maximum magnitude of the waveform. 
The denominator uses the root-mean-square (RMS) as an estimator for the noise power in consecutive, non-overlapping time intervals $n_j$ of equal length.
The median is chosen to avoid the impact of occasional outliers, such as human-made or natural radio pulses, as well as the air-shower pulse itself, on the estimated noise power.
Although counterintuitive, it is common in air-shower radio detection that pure background traces yield average SNR values well above 1. In our case, only SNR values $\gg 10$ correspond to radio pulses that are clearly distinguishable from noise by eye.
Figure~\ref{fig:snr_dist} shows the SNR distribution for both polarizations of antenna\,1.

\section{Network Performance}\label{sec:network_results}
This section describes the performance evaluation of the networks using the test dataset. 
For the \classifier\, evaluation is straightforward, based on correct and false classifications. For the \denoiser\, physics motivated metrics are used, specifically focusing on the accurate reconstruction of the radio pulse's peak time and power, as these are critical parameters for air-shower measurements.

\subsection{Classifier}

\begin{figure*}[hbt!]
\centering
\includegraphics[width=0.48\textwidth]{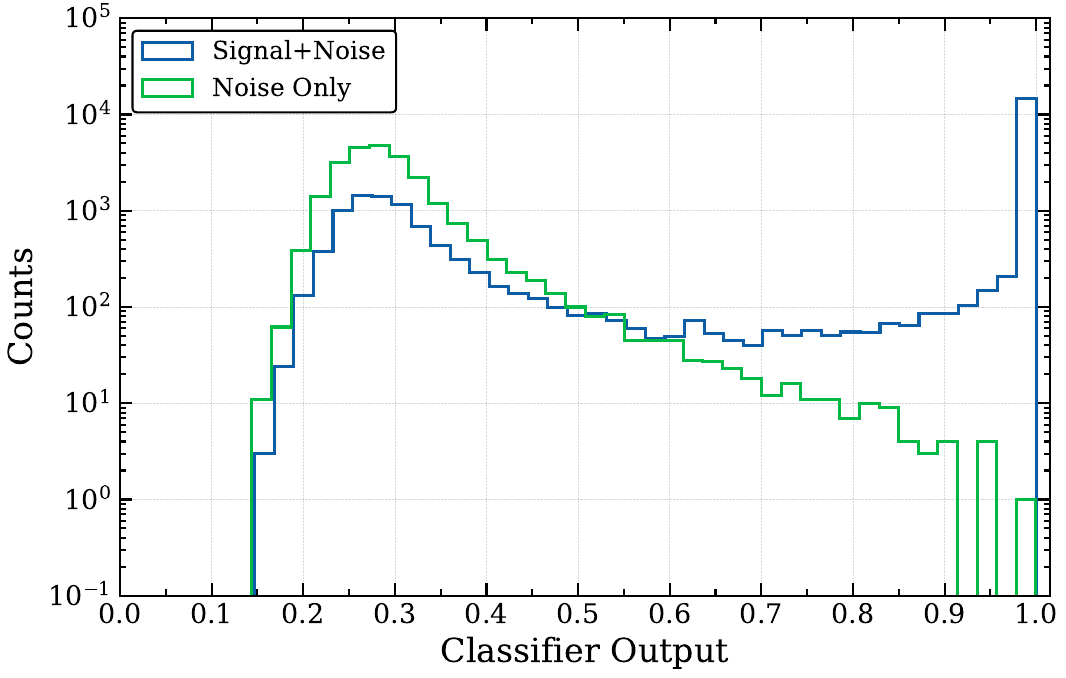}
\includegraphics[width=0.48\textwidth]{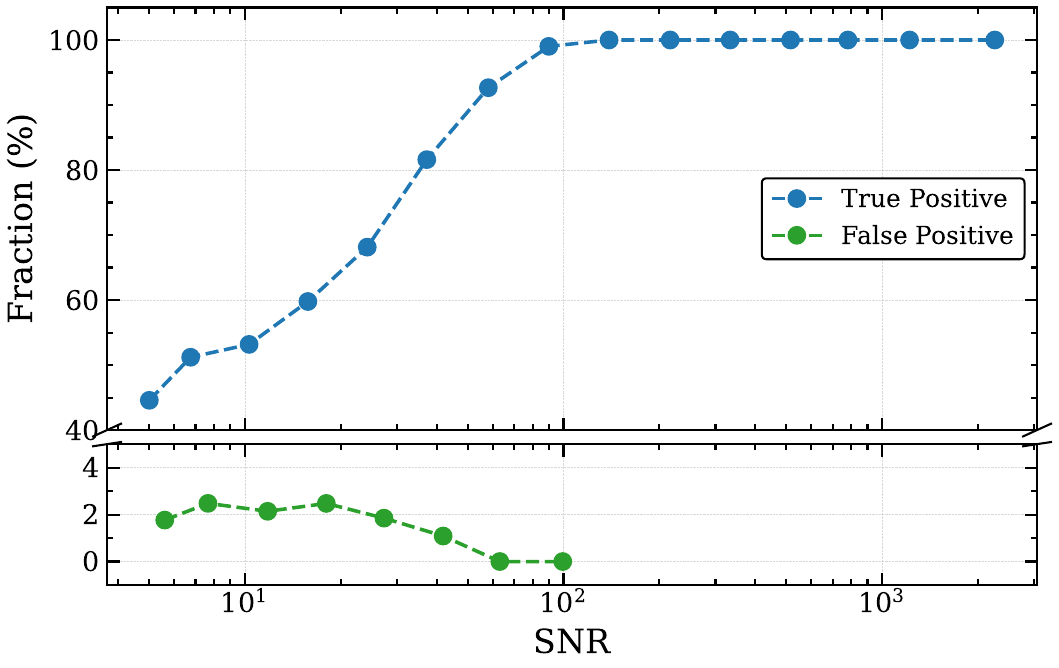}
\caption{Left: classifier output obtained for both the noisy waveform data (blue) and for the noise-only waveform data (green). Right: true positive and false positive rate as a function of SNR, obtained after applying a threshold cut of 0.5 on the classifier output. Note that this threshold is used only to illustrate performance on the test set. The y‑axis is broken to emphasize low false‑positive fractions. The false-positive curve stops at SNR=100, as almost none of the noise-only waveforms in the data sample exceeds that value (cf.~Fig.\ref{fig:snr_dist}).}
\label{fig:classifier}
\end{figure*}

The \classifier\ is trained on a mix of noisy (signal + noise) and noise-only data, with the integer values $1$ and $0$ used as labels, respectively. Each input sample, consisting of two polarization waveforms, produces a single output value between 0 and 1. The left panel of Fig.~\ref{fig:classifier} shows the distribution of classifier outputs for the test set. 
While the network effectively pushes signal and background events towards outputs of 1 and 0 respectively, overlap remains, especially for scores below 0.5, where classification becomes more challenging.

To illustrate the performance of the \classifier, a threshold of 0.5 is applied, classifying samples with outputs $\geq$\,0.5 as signal-like and others as background-like. 
This results in an overall false positive rate (FPR) of 2.2\%, i.e., background traces mis-classified as air-shower pulses.

In contrast to many other applications of classifiers, the concept of a true-positive rate (TPR) is not meaningful for externally triggered radio enhancements. 
All air-shower events are known to contain some radio signal (as by definition also all events in our noisy dataset), but often this radio signal is negligilbe and too weak to be detected.
This depends on many parameters, such as the energy of the air shower and its geometry relative to the antenna. 
In a naive approach counting all events, the TPR would thus drastically depend on how many air showers with undetectable weak signal would be included in the sample, and any definition of a detectability threshold would be arbitrary.

We have thus divided the test dataset into SNR bins, and determined the TPR and FPR for each bin. Rates are reported as percentages, with the TPR defined as the number of correctly identified signals divided by the total number of noisy waveforms.
The right panel of Fig.~\ref{fig:classifier} shows TPR and FPR as a function of SNR.
A TPR above 80\% was achieved for SNR $>$40, while the FPR remained below 5\% across all SNR values.

In practice, the particular situation of the experiment can guide how the \classifier\ is used. For example, in our case, the IceTop array at the same location allows an additional cross-check, enabling us to tolerate a few percent FPR at the \classifier\ stage.
Moreover, to identify radio events from air showers, both the \classifier\ score and the amplitude of the denoised pulse are considered in a combined selection criterion (see Section.~\ref{sec:application} for more details).

\subsection{Denoiser}

\begin{figure*}[hbt!]
\centering
\includegraphics[width=0.94\textwidth]{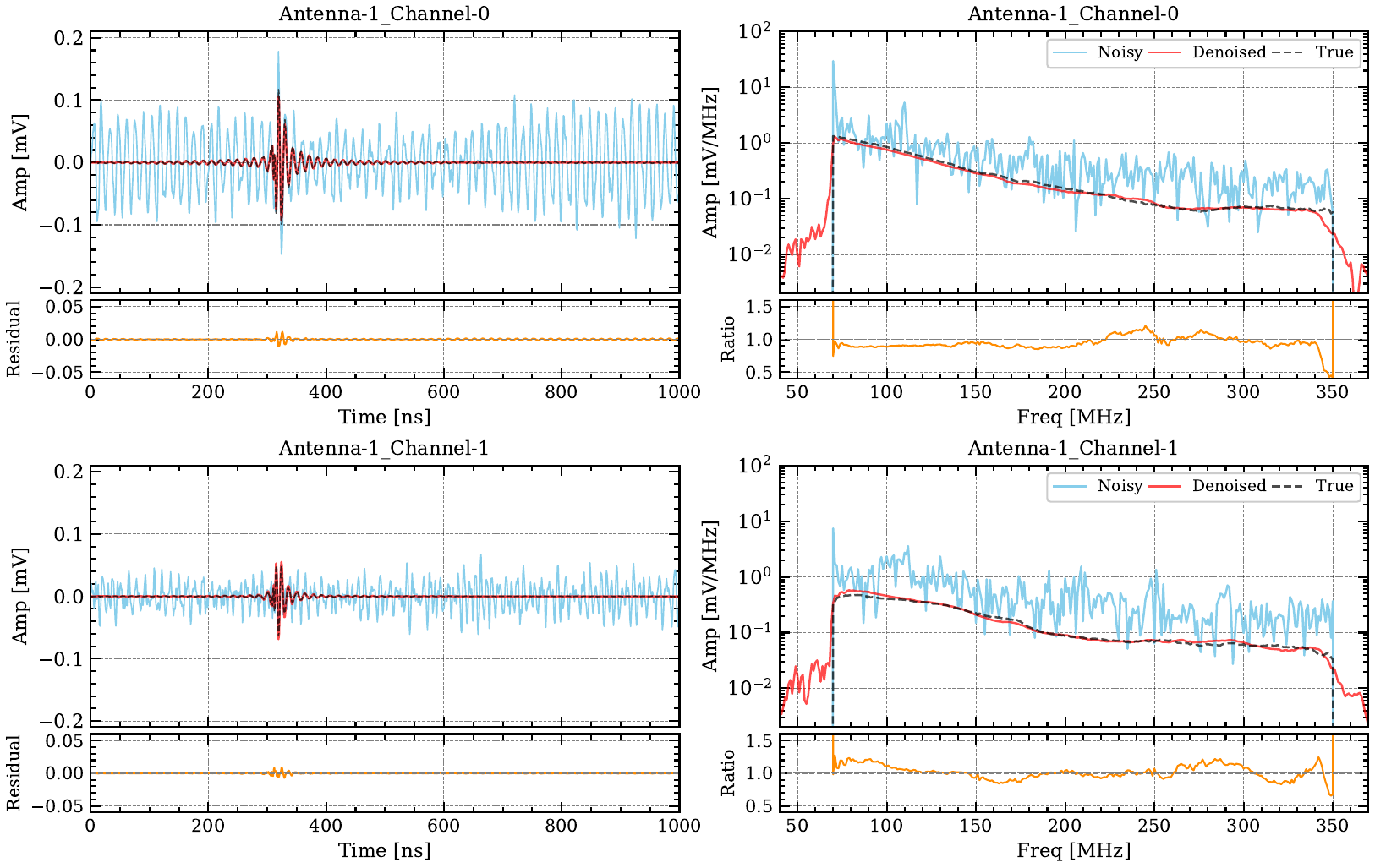}
\caption{Example of a denoised waveform, together with the corresponding noisy input and true waveform (CoREAS simulation including the antenna response), shown for the two polarization channels of the antenna. The frequency spectra of the waveforms are displayed in the right panels. The lower subpanels present the difference between the denoised and true waveforms in the time domain, and the ratio between the denoised and true spectra in the frequency domain.}
\label{fig:denoising_example}
\end{figure*}
 
Following training, the \denoiser\ was also evaluated using the test dataset. 
Noisy traces identified as signal-like (i.e., with classifier scores $\geq$ 0.5) were passed through the \denoiser\ for cleaning. 
An example of a successful denoising result is shown in Fig.~\ref{fig:denoising_example}, where the radio pulse shape and timing are  recovered in both channels and the background is strongly suppressed inside of the design band of $70-350\,$MHz (differences outside of the design band do not matter for air-shower analyses, as these usually filter to a sub-band fully contained in the design band).

\begin{figure*}[hbt!]
\centering
\includegraphics[width=0.49\textwidth]{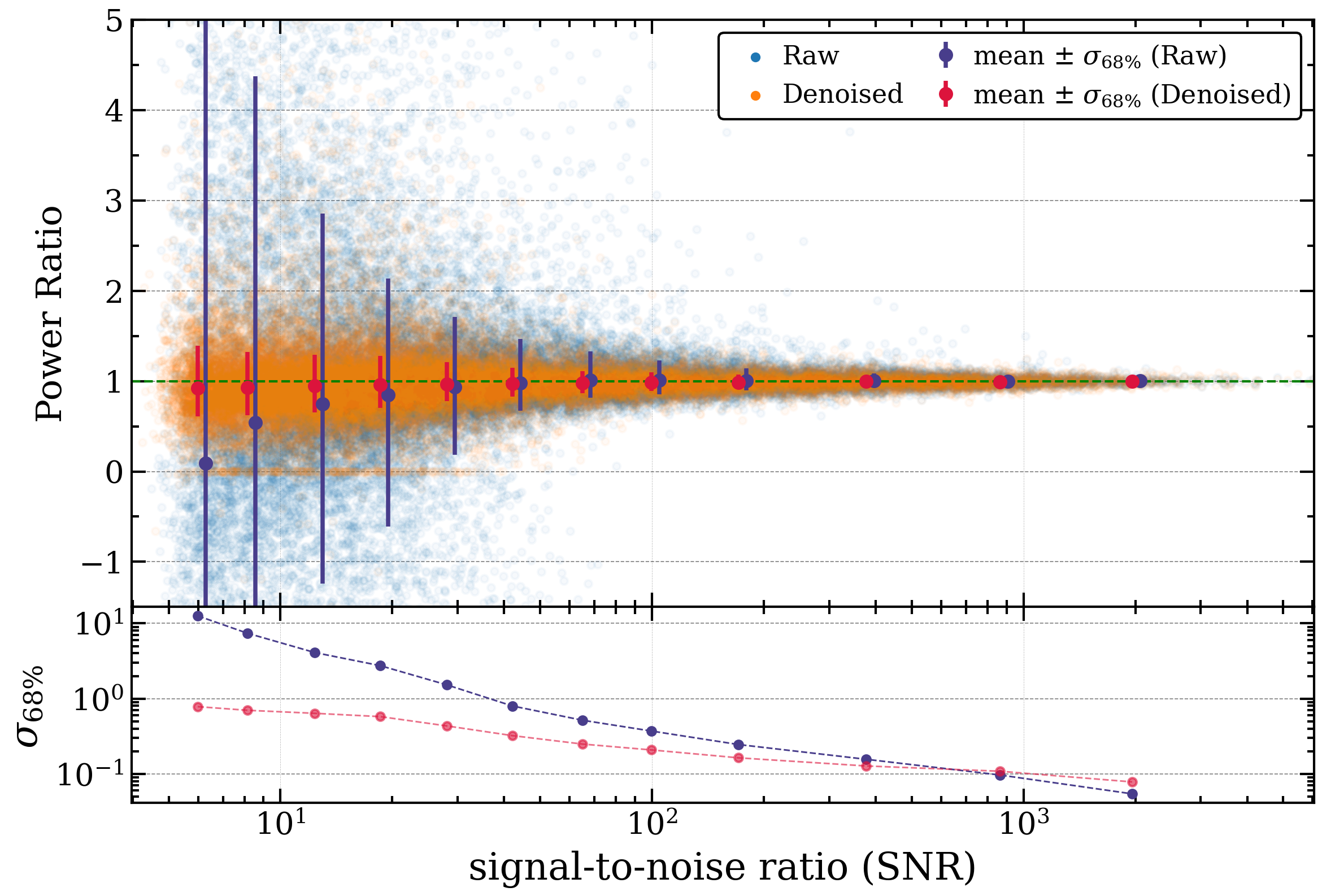}
\includegraphics[width=0.49\textwidth]{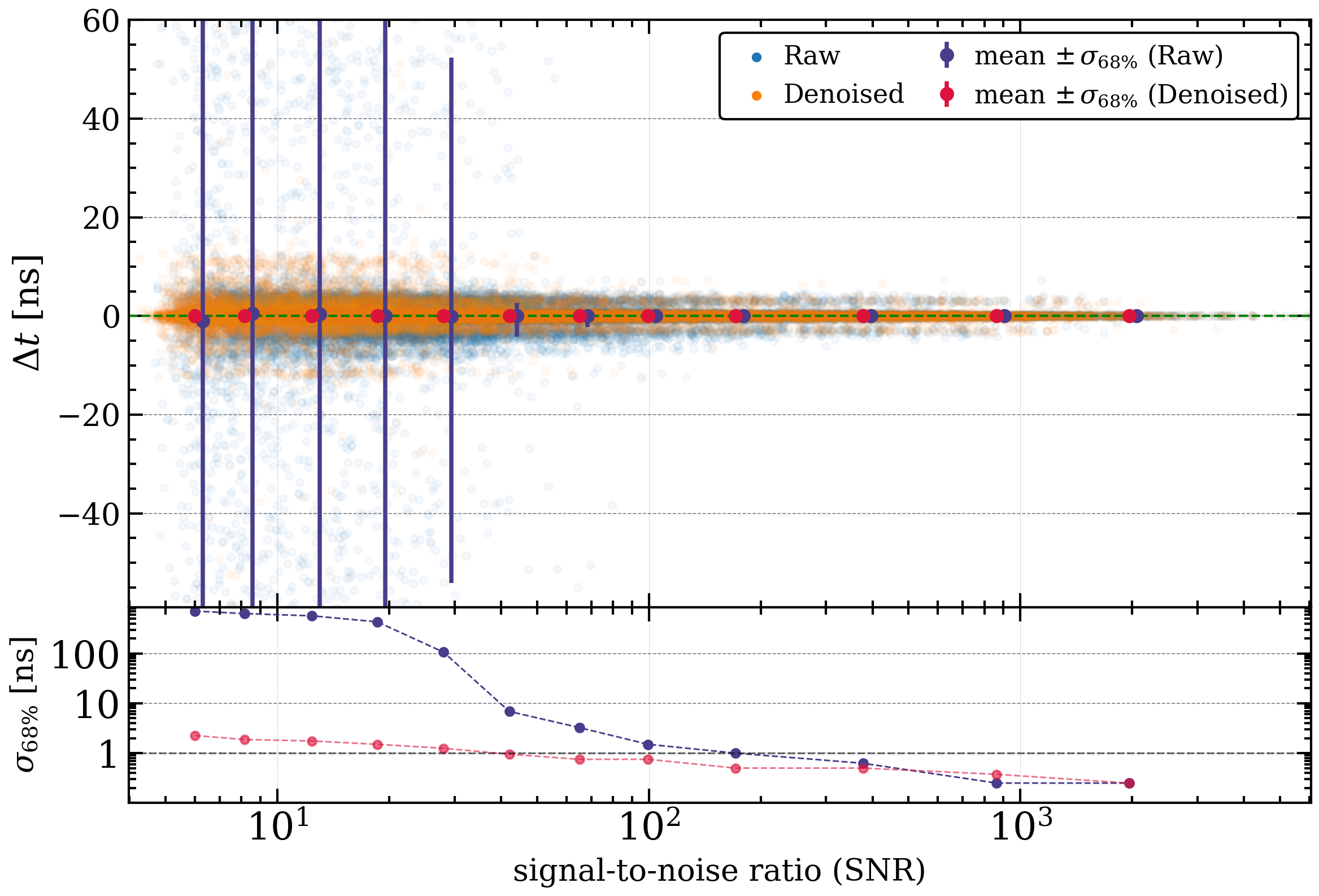}
\caption{Left: Power ratio as a function of SNR for both waveforms without applying denoiser (blue distribution) and after applying the denoiser (orange distribution). Right: Peak time difference over SNR. The navy and red dots are the mean values for (not-)denoised waveforms with the error bars representing 68\% containment, which is also shown in the bottom plot for better visibility.}
\label{fig:denoiser}
\end{figure*}

To quantify the \denoiser's performance with quantities relevant for physics analyses, two accuracy metrics were devised: power ratio and peak time difference. 
These metrics compare raw (non-denoised) and denoised waveforms.

The power ratio is calculated by measuring the power within a 50\,ns signal window ($P_S$) centered around the simulated pulse position, and a 400\,ns non-overlapping noise window ($P_N$). The power ratio is then given by:

\begin{equation}
   \text{Power Ratio}~=~ \frac{ \left[ P_S -P_N \right]_{\rm measured}}{\left[ P_S -P_N \right]_{\rm true}}~.
\end{equation}

In the ideal case, the power ratio is $1$, indicating that the calculated power matches the true radio pulse power. Negative values can occur for small signals when the power measured in the background window exceeds the power in the signal window.

The peak time difference, $\Delta t$ is computed by locating the peak of the waveform’s Hilbert envelope, i.e., the magnitude of its analytic signal that represents the instantaneous amplitude, and subtracting the true pulse time:

\begin{equation}
   \Delta t = T_{\rm measured} - T_{\rm true}~,
\end{equation}
where $T_{\rm measured}$ is the peak time (from either the raw or denoised traces) and $T_{\rm true}$ is the simulated pulse time.
Ideally, $\Delta t =0$, meaning the estimated time matches the true radio pulse time within the trace.

The power ratio as a function of SNR is presented in the left panel of Fig.~\ref{fig:denoiser}. At low to intermediate SNR values ($\leq$ 100), the \denoiser\ reduces both bias (deviation from an ideal power ratio of $1$) and spread, improving the estimation of the pulse power. 
Near-zero power ratios in the denoised waveforms arise when the \denoiser\ fails to recognize a pulse and mistakenly removes both pulse and background.
At high SNR, both raw and denoised traces yield similarly accurate results. 

The right panel of Fig.~\ref{fig:denoiser} shows the peak time difference as a function of SNR. 
Similar to the power ratio, the \denoiser\ notably improves time accuracy at low and intermediate SNR. At high SNR, there is only a slight difference between denoised and non-denoised cases, with typical time offsets $\leq 1$ns. 
The large scatter in raw values is due to background fluctuations being misidentified as the pulse, since the highest peak across the entire trace is taken as the pulse peak time. 
The horizontal banded pattern seen in the distribution is caused by the algorithm sometimes detecting different oscillations of the waveform for peak time, effectively selecting one oscillation peak over another \cite{Rehman:2023jme}.

In summary, the \denoiser\ significantly enhances the accuracy of pulse power and peak time at low to intermediate SNR, while at high SNR the impact of background is anyway low, and the accuracy is fully sufficient in either case.


\section{Application of Networks to Measured Air Showers}\label{sec:application}

\begin{figure*}[hbt!]
\centering
\includegraphics[width=0.77\textwidth]{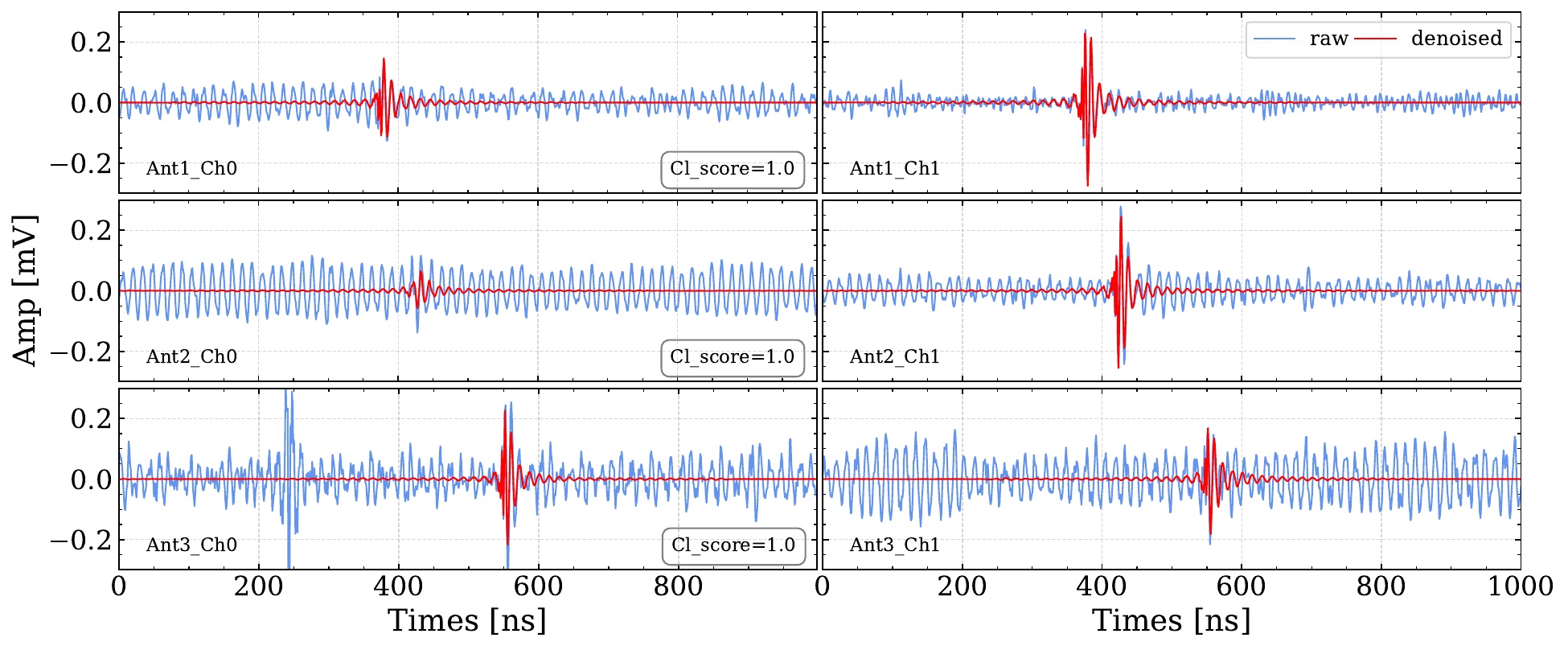}
\caption{Measured waveforms for all six channels from the three antennas for an air-shower event triggered by the scintillation panels. 
The raw waveforms, utilized as input for the networks, are depicted in blue. 
The cleaned waveforms, obtained through the denoising process, are presented in red. The output of the Classifier, represented as the Cl\_score, is displayed as text for each of the three antennas. In this particular example, all three antennas yield a Cl\_score of 1.0, representing an ideal classification that is not always achieved.
}
\label{fig:example}
\end{figure*}

After the testing phase, the networks were applied to air-shower events simultaneously measured by the three surface antennas of the prototype station, triggered by the scintillation panels.
Figure~\ref{fig:example} presents an example event: six waveforms corresponding to the two polarization channels of the three antennas.
The blue curves show the measured, preprocessed waveforms (see Sec. \ref{sec:dataPrep}), which serve as inputs to the networks. The red curves depict the output from the \denoiser\, and \classifier\ scores (Cl\_score) for each antenna are shown in the text box.
In this example, all three antennas received the maximum Cl\_score, and the \denoiser\ successfully suppressed background noise, even in channels with significant interferences. 
Each antenna recorded a strong signal in at least one polarization channel, while signals in the other channels were weak or comparable to the background.  
Additionally, in channel 0 of antenna 3, an artifact taller than the true radio signal appears around 250\,ns; the network correctly identified and removed this feature, which standard methods looking for the highest peak in the trace would have misidentified as a signal. 
The resulting directional reconstruction for this event differs by only $2.6^\circ$ from the IceTop reconstruction of the same air shower, as expected when the network correctly identified the air-shower radio signal (see next section).

However, such optimal results are not always guaranteed. Most scintillator-triggered events have radio signals below the detection threshold.
To filter out poorly denoised events without a clear radio signal, a set of quality cuts was developed. 
The first cut uses the output of both the \classifier\ and \denoiser\ to retain waveforms likely containing an air-shower radio pulse. 

\begin{figure}[hbt!]
\centering
\includegraphics[width=0.49\textwidth]{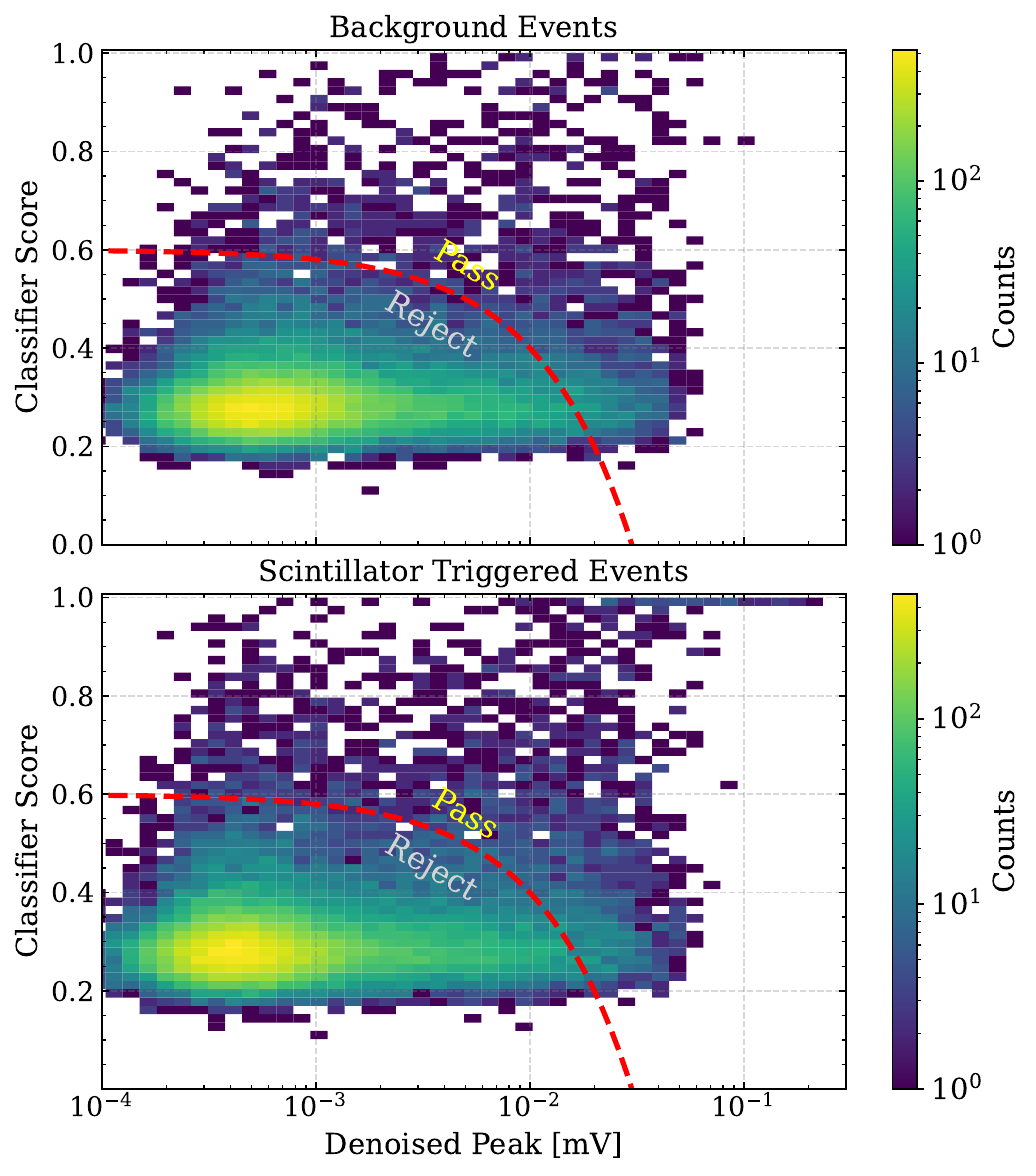}
\caption{Top: Distribution of the \classifier\ output (y-axis) versus the maximum amplitude of the denoised waveform (x-axis) for pure background events recorded with the fixed-rate trigger (FRT). The red dashed line indicates the quality cut used to reject 95\,\% of background waveforms; only waveforms above this line are considered as passing. Bottom: Same distribution for scintillator-triggered air-shower events. A distinct population in the upper right region indicates clear air-shower signals.}
\label{fig:quality_cuts}
\end{figure}

The cut values were determined by applying the networks to background waveforms from FRT events as shown in the top panel of Fig.~\ref{fig:quality_cuts}. The plot shows the distribution of classifier scores (y-axis) versus the absolute peak amplitude of the denoised waveform (x-axis).
The threshold is defined by a line $y = mx + c$, shown as a red dashed line, with the slope ($m = -20$) and intercept $(c = 0.6)$ tuned to achieve 95\,\% rejection of of pure background waveforms per antenna, comparable to reference methods based on a threshold in SNR~\cite{Abbasi:2024ws}.
By this choice of a cut combining the classifier output and the pulse amplitude, we identify air-shower radio pulses of low amplitude without the risk of losing air-shower events with high-amplitude radio pulses (using a flat cut of 0.6 in \classifier~score would result in an approximately 3-fold increase in detected air-shower events over the reference SNR method, while the combined cut results in a 5-fold increase, as shown later).
For events surviving this cut across all three antennas, a subsequent cut removes those whose reconstructed radio arrival direction is inconsistent with IceTop.

The bottom panel of Fig.~\ref{fig:quality_cuts} displays the same network outputs, but for scintillator-triggered events. As expected, most outputs cluster in the lower-left corner of the plot, indicating that the networks categorize them as background-like. 
This is consistent with expectations, as the scintillator panels have lower detection thresholds and trigger on lower-energy air showers with little or no detectable radio signal. 
Nevertheless, a small cluster of events in the top-right of the plot indicates confident detections of air-shower radio signals.

The distribution of the \classifier\ scores and denoised peak amplitudes illustrates the flexibility of the selection strategy: the decision boundary can be adjusted depending on whether a particular analysis prioritizes purity (requiring higher scores/amplitudes) or efficiency (accepting lower values).
There generally is no need for perfect classification, as the IceTop measurement of the same air showers is always available for cross-check.
The choice used here is motivated by achieving $95\,\%$ background rejection per antenna, to compare the CNNs with the reference SNR method.

\begin{figure}[hbt!]
\centering
\includegraphics[width=0.48\textwidth]{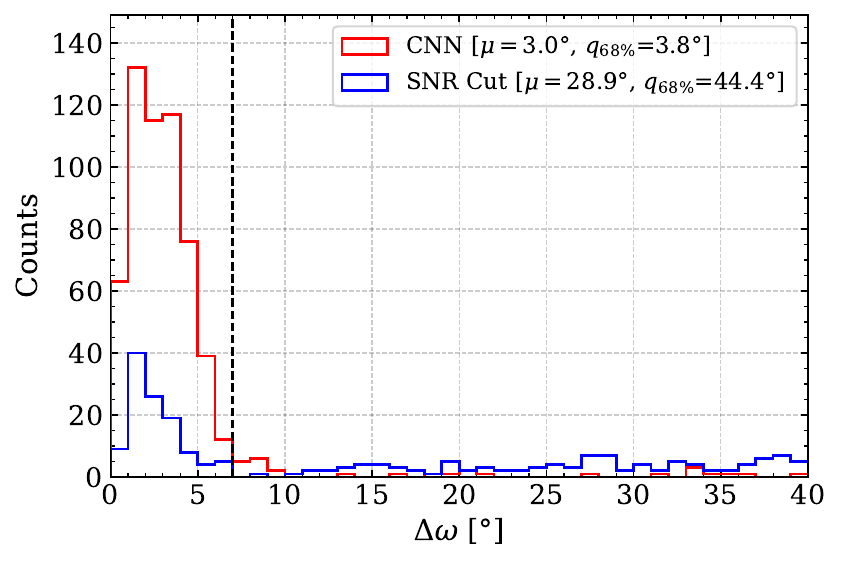}
\caption{Opening angle between the radio-reconstructed and IceTop-reconstructed directions for scintillator-triggered air-shower events identified using the CNN method and an SNR method (see Sec.~\ref{sec:comparison}). The dashed line at $7^\circ$ indicates the quality cut applied to select well-reconstructed air-shower events. The mean values of the open angle for events within this quality cut are $3.5^\circ$ for the CNN and $3.0^\circ$ for the SNR method.}
\label{fig:TradVsCNN}
\end{figure}

\subsection{Reconstructed Air-Shower Events and their Angular Distributions}\label{ss:angle}

For the successful reconstruction of the air-shower arrival direction, at least one channel from each of the three antennas must pass the quality cuts described in the previous section. When this condition is met, the denoised waveforms are used for direction reconstruction. From each antenna, the channel with the largest peak amplitude is selected to extract the pulse arrival time.
Because the prototype station only features three antennas, the arrival direction is reconstructed assuming a simple plane wavefront.

During a 106-day search period, a total of 227,613 scintillator-triggered events were recorded and analyzed using the CNN-based pipeline.
Among these, 608 events passed the CNN quality cuts and were considered for arrival direction reconstruction. The reconstructed directions from radio were then compared to those obtained from IceTop. Fig.~\ref{fig:TradVsCNN} shows the distribution of opening angles ($\Delta \omega$) between the two reconstructions. As seen in the zoomed-in view, most events cluster within a few degrees of the IceTop direction, indicating accurate identification of the radio pulse in all three antennas.

A small subset of events, however, shows large opening angles, forming a tail in the distribution. These events typically result from incorrect identification of the radio pulse in at least one antenna, for example, when an artifact or background fluctuation is mistaken for a true signal.

To exclude such cases, an additional data-driven quality cut is applied.
Events with $\Delta \omega$ $\geq$ 7\Deg~are removed, where the distribution shows a significant drop. 
As the main purpose of the radio enhancement is to increase the measurement accuracy of IceTop for cosmic-ray analyses, this IceTop direction reconstruction is available for all air-shower events of interest.
The choice of $\Delta \omega$ threshold can be adjusted based on specific science analysis requirements for purity and efficiency.
This cut at 7\Deg removes about $9\,\%$ of the events found with the CNNs, leaving 554 candidate radio events.

\begin{figure}[hbt!]
\centering
\includegraphics[width=0.37\textwidth]{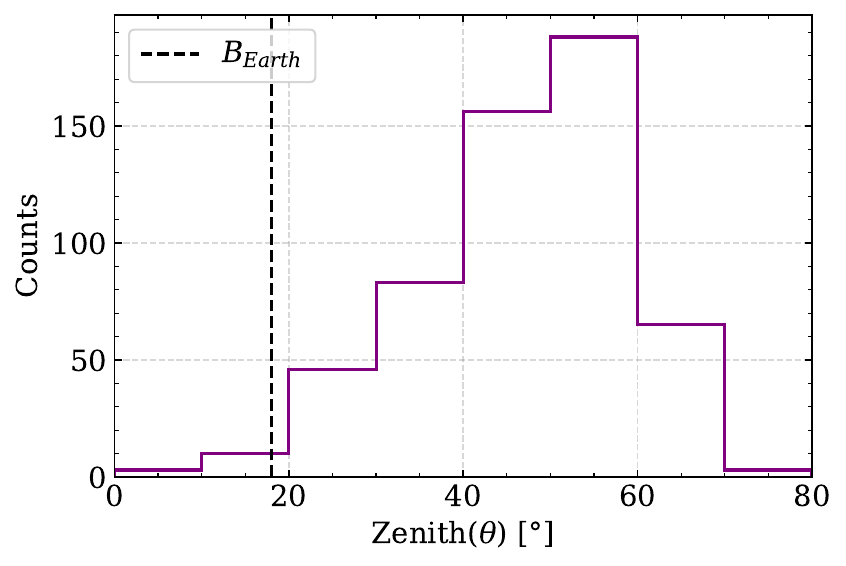}
\includegraphics[width=0.37\textwidth]{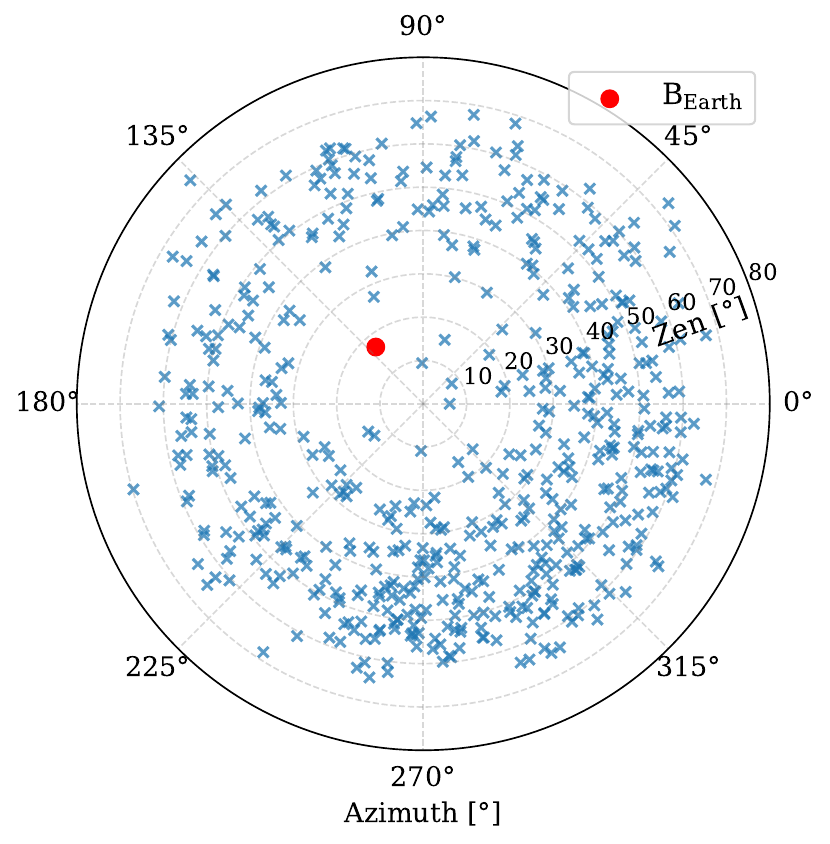}
\caption{Zenith angles (top) and Arrival directions (bottom) of the candidate radio events. The black dashed line in the upper plot and the red-filled circle in the lower plot represent the local geomagnetic field at the IceTop location. }
\label{fig:zenazi}
\end{figure}

Since the strength of the radio emission depends on the geomagnetic Lorentz force, we expect a suppression of events aligned with the geomagnetic field and an enhancement at larger geomagnetic angles (the angle between the shower axis and the local geomagnetic field).
This expectation is reflected in the zenith distribution, which shows a concentration of events at intermediate zenith angles.
The decline above 60\Deg\ is due to the reduced efficiency of the scintillator trigger at large angles. 
Additionally, we require coincidence with IceTop, and the IceTop array also shows decreasing sensitivity at high zenith angles \cite{Paudel:2024wtm}. 
Again, consistent with the expectation of geomagnetic radio emission, the azimuth and polar distributions also show an excess of events in the direction opposite to the geomagnetic field.

\subsection{Core Distribution}\label{ss:core}

\begin{figure}[hbt!]
\centering
\includegraphics[width=0.45\textwidth]{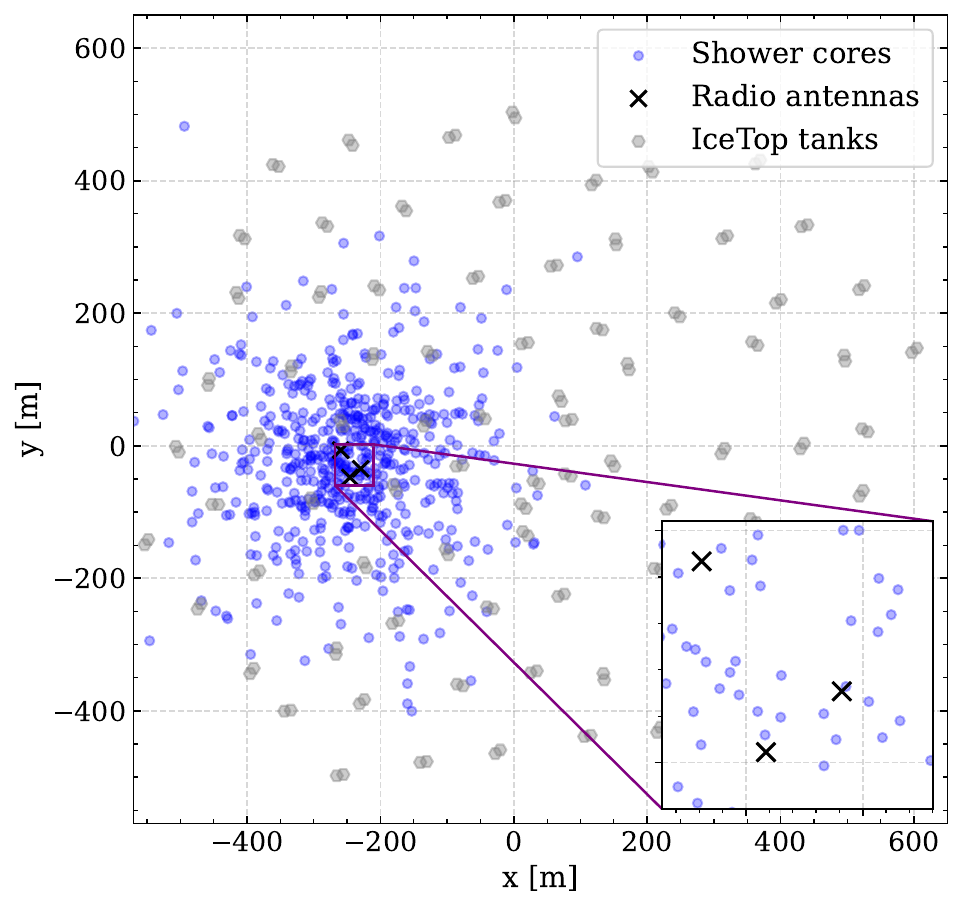}
\caption{IceTop core distribution of the air-shower events detected with the radio antennas. Shower core positions are marked with blue circles, antenna locations are indicated by black ``X''s, and IceTop tanks are represented by grey circles.}
\label{fig:cores}
\end{figure}

The shower core positions, defined as the impact point of the shower axis on the ground, were reconstructed using IceTop data and are shown in Fig.~\ref{fig:cores}, along with the locations of the detectors. The figure demonstrates that most candidate events have core positions close to the prototype station. However, several events were detected with cores located more than $200\,$m away from the antennas. 
These correspond to inclined showers, which produce elongated radio footprints due to their geometry, allowing the radio signal to reach the array from a greater distance.

\begin{figure}[hbt!]
\centering
\includegraphics[width=0.44\textwidth]{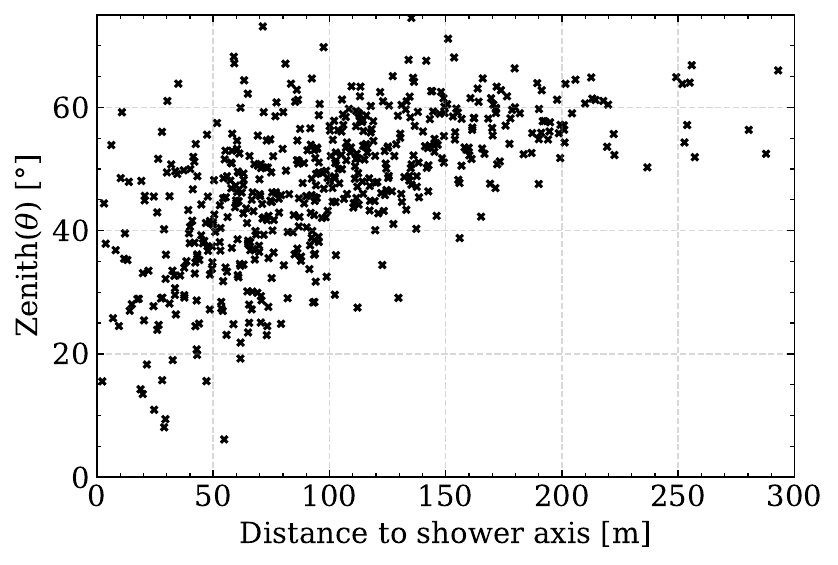}
\includegraphics[width=0.44\textwidth]{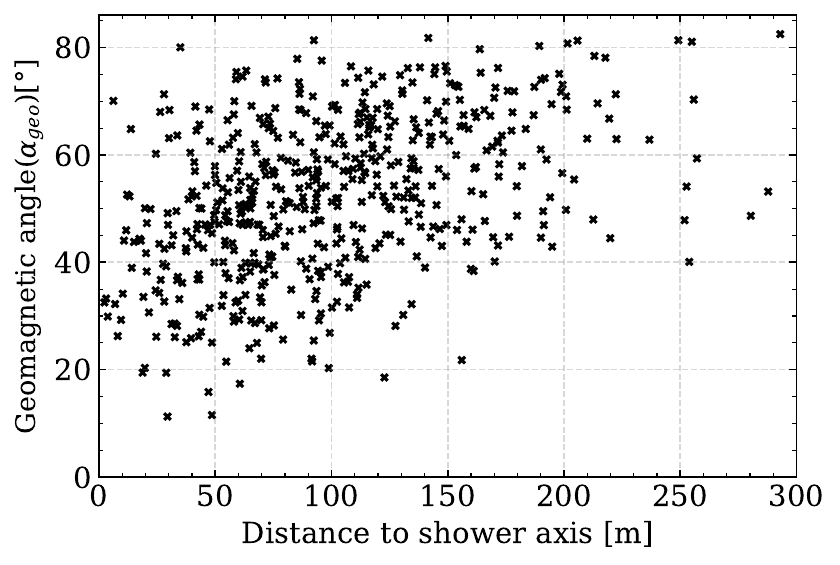}

\caption{The distance of the shower core to the center of the prototype station is calculated in the shower plane and plotted over the zenith angle (top) and geomagnetic angle (bottom). The events that are far from the station are also the ones with larger zenith and geomagnetic angles, hence a large radio emission from these events is expected.}
\label{fig:coredist_vs_angles}
\end{figure}

Figure~\ref{fig:coredist_vs_angles} shows the distance between the prototype station center and the reconstructed shower axis, plotted as a function of the zenith angle (top panel) and the geomagnetic angle (bottom panel). The top panel illustrates that events farther from the array have larger zenith angles, some even exceeding 60\Deg, consistent with expectations for inclined showers that produce larger radio footprints.

The bottom panel further shows that these events also have larger geomagnetic angles, which indicates stronger geomagnetic radio emissions. 
Overall, these trends provide strong evidence that the CNNs correctly identified radio signals from cosmic-ray air showers.

\section{Comparison with an SNR method}\label{sec:comparison}

To realistically assess the CNN performance on experimental data, we compare its results with those obtained from a standard SNR method that combines narrow-band radio frequency interference (RFI) suppression via digital filtering and a SNR threshold cut.
Variants of this standard method have been used by several radio arrays for air-shower detection (see e.g.~\cite{PierreAuger:2016vya,Bezyazeekov:2018yjw}), and have also been applied to data from the prototype station at IceTop~\cite{IceCube:2021epf}.

In this SNR method, a bandpass filter is applied to limit the signal to the $100 - 230\,$ MHz range, which has been optimized using CoREAS simulations to enhance the SNR for typical air-shower signals and is consistent with earlier simulation studies~\cite{BalagopalV:2017aan}.
To further suppress narrow-band RFI within this band, frequency-dependent weights are applied~\cite{IceCube:2021epf}.
Note that, for the CNN method, the full band of $70 - 350\,$MHz is used and does not require explicit RFI suppression.
After filtering, a SNR threshold is applied, chosen to reject 95\,\% of the background waveforms (FRT events) in each channel.

All scintillator-triggered events within the search period are considered for which at least one channel from each of the three antennas exceeds the SNR threshold. 
These selected events are then reconstructed using a plane-wave front model, as described earlier for the CNN method.

Using this SNR method, 341 events were successfully reconstructed. The opening angle between their radio-reconstructed directions and the corresponding IceTop directions is shown in Fig.~\ref{fig:TradVsCNN} (blue). 
Similar to the CNN-based results, many events are reconstructed within a few degrees of the IceTop direction, however, the distribution features a more pronounced tail at large opening angles.
Applying the same quality cut of $\Delta \omega \leq 7$\Deg, yields 111 candidate events, approximately one-fifth the number identified using the CNN method.
Despite the much larger number of CNN events passing the quality cut, which are mostly at lower energy and therefore at lower SNR than the events identified by the SNR method, the mean opening angle for those events  deteriorates only slightly from $3.0^\circ$ to $3.5^\circ$.

\begin{figure*}[hbt!]
\centering
\includegraphics[width=0.47\textwidth]{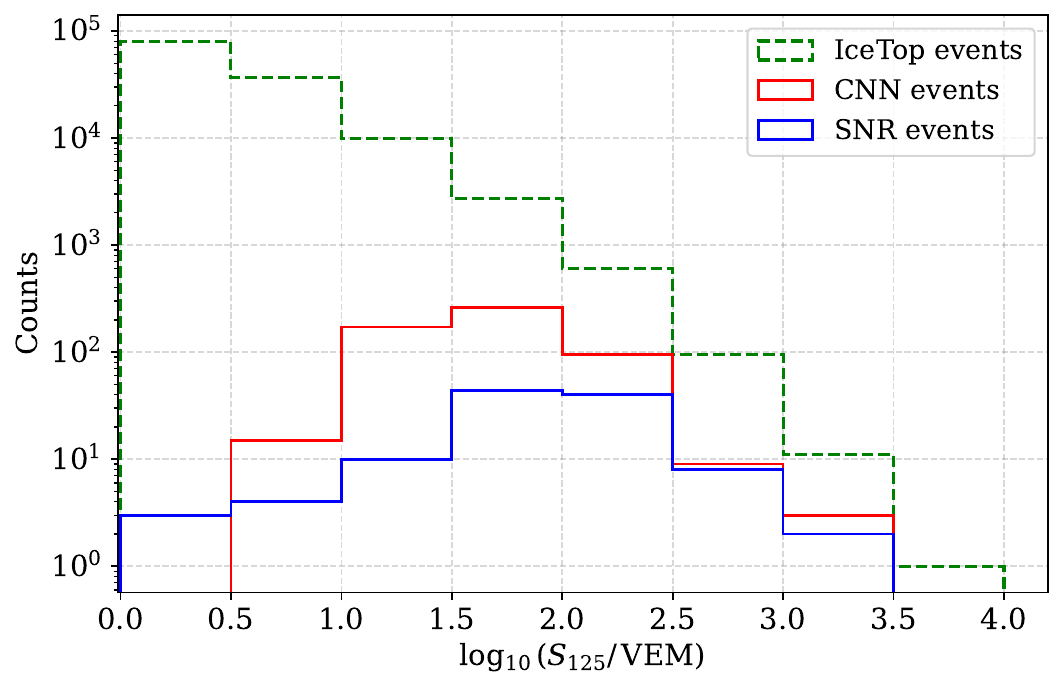}
\includegraphics[width=0.47\textwidth]{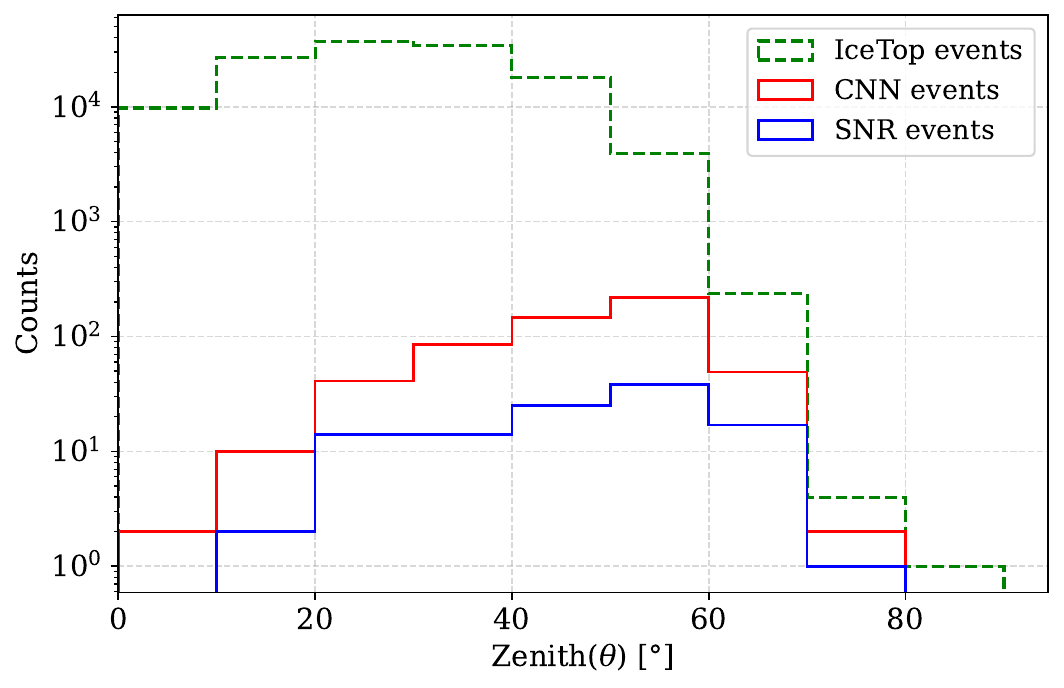}
\includegraphics[width=0.47\textwidth]{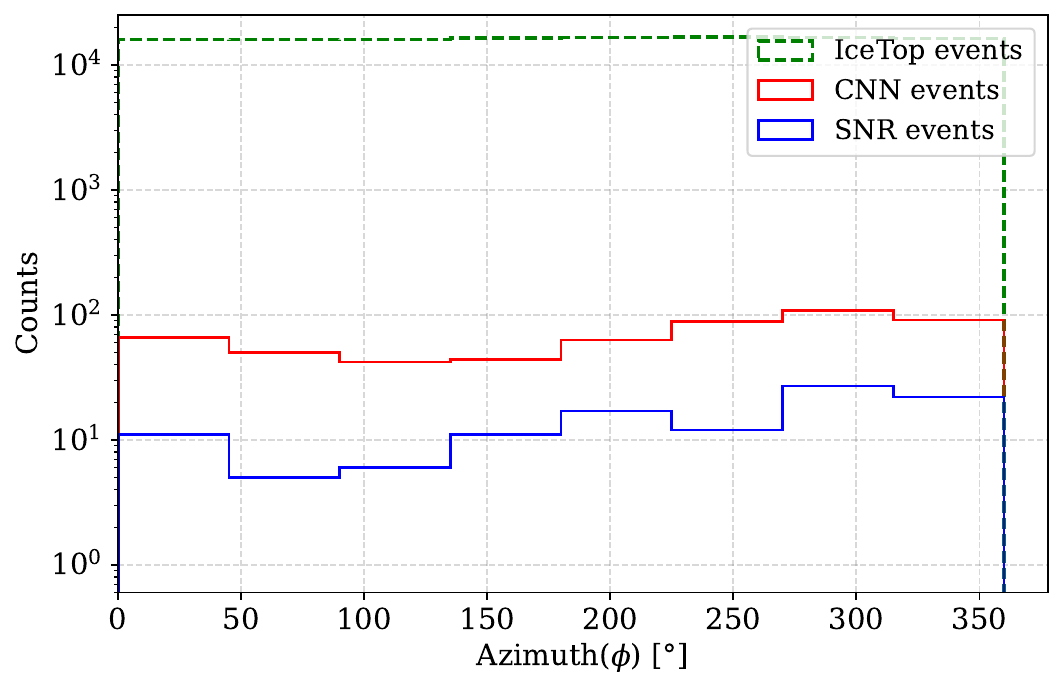}
\includegraphics[width=0.47\textwidth]{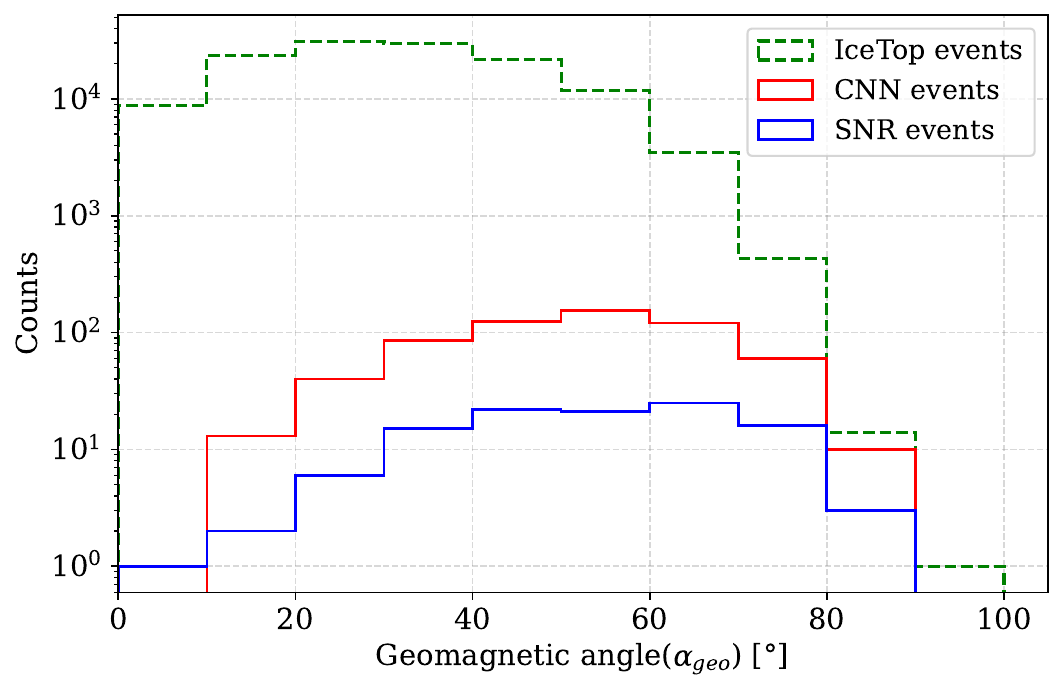}
\caption{Distributions of air-shower events identified with the CNN (red) and SNR (blue) methods. The top-left panel shows the distribution of \logE, i.e., the IceTop signal strength at $125\,$m from the shower axis in units of vertical equivalent muons (VEM). The other panels show the distributions of zenith angle (top-right), azimuth angle (bottom-left), and geomagnetic angle (bottom-right). For reference, all IceTop events with a scintillator trigger are shown as a green dashed line.}
\label{fig:s125AndZen}
\end{figure*}

Figure~\ref{fig:s125AndZen} (top-left) compares the \logE\ distributions for candidate events identified by both methods, along with all scintillator-triggered IceTop events. 
The \logE\ parameter (in units of Vertical Equivalent Muons (VEM)) is used as an energy proxy for IceTop, however, the energy reconstruction is only optimized and calibrated for small zenith angles ($\theta < 30^\circ$) \cite{IceCube:2012nn, IceCube:2023nyj}.
Most high-energy events (on the right-hand side of the \logE\ distribution) were identified by both methods; however, the CNN method recovers significantly more lower-energy events, which are missed by the SNR method. 
Only 8 of the 111 events found by the SNR method were missed by the CNN method. 
These 8 events are all at low energy \logE$ < 1$, and may therefore be false-positive detections by the SNR method.
This would be consistent with their non-identification by the CNN method.
As discussed in the next section, the false-positive rate is indeed expected much larger for the SNR method than for the CNN method.

The zenith-angle (top-right), azimuth-angle (bottom-left), and geomagnetic angle (bottom-right) distributions in Fig.~\ref{fig:s125AndZen} further demonstrate that the increase in the number of events identified by the CNN method is consistent across all angular ranges.

\subsection{Estimation of the Rate of False Positive Events}
\label{ssec:bacground_estimation}

Even in an externally triggered radio setup, like the IceTop enhancement, false positive detections of radio pulses may negatively affect physics analyses.
Because of the IceTop information available for each air shower, the tolerable false-positive rate (FPR) for the additional radio measurement of the same air showers will be higher than in self-triggered radio instruments, typically of at least per mill level, to be determined for each cosmic-ray analysis individually. 

The vice-versa concept of a true-positive rate (TPR) is less useful for our application, as the radio readout is triggered by the scintillation panels. 
The absolute value of the TPR would thus depend critically on the trigger threshold, which is purposefully chosen such that the radio signal is well below the noise level for most triggered air showers. 
As shown in the previous section, we correctly detect the air-shower radio signal in about five times as many events than with the SNR approach, which corresponds to a five-time relative increase of the TPR.
Here, we focus on determining how the CNNs impact the FPR compared to the SNR method, i.e., for how many of the air-shower events background would be mistaken for the radio signal. 

The selection cuts were designed to reject $95\,\%$ of background waveforms per antenna. Consequently, some background events will pass this cut in all three antennas simultaneously, and a few of those events may coincidentally yield an arrival direction consistent with IceTop.
Because the background across multiple antennas is correlated, it is insufficient to estimate the rate of false positive events by simply multiplying individual probabilities. 
Therefore, to obtain a realistic estimate, we evaluate the FPR for both methods by applying the full selection criteria to FRT events and subsequently reconstructing the arrival direction using the same procedure used for actual air-shower candidates. 

\begin{table*}[ht]
   \centering
   \caption{Summary of false positive estimates for the SNR and CNN methods. Of all FRT events, 321 passed the SNR selection and were reconstructed, compared to only 8 for the CNN method. After the $\Delta\omega$ cut, the expected number of false positives is 3.7 for SNR and 0.08 for CNN. These values are obtained by multiplying the final false-positive fractions (last row) with the total number of scintillator-triggered events during the search period (227,613 events).}
   
    \renewcommand{\arraystretch}{1.5}
    \begin{tabular}{|l|c|c|c|}
        \hline
        \textbf{} & \textbf{Total events} & \textbf{Passing Fraction} & \textbf{Passing Fraction} \\ 
        \textbf{} & & \textbf{SNR cuts} & \textbf{CNN method} \\ 
        \hline
        Fixed-Rate Triggered (FRT) & 336070 & $9.6 \times 10^{-4}$ ~ (321 events) & $2.4 \times 10^{-5}$ ~ (8 events) \\
        \hline
        Mean fraction of IceTop events within $7^\circ$ & & 0.017 & 0.014 \\
        \hline \hline
        Resulting false-positive fraction & & $1.6 \times 10^{-5}$ & $3.3 \times 10^{-7}$ \\
        \hline
    \end{tabular}
    
    \label{tab:FPcomparison}
\end{table*}

During the 106-day search period, a total of 336,070 FRT events were recorded. Of these, 321 events passed the SNR selection in all three antennas and were successfully reconstructed. In comparison, only 8 FRT events passed the CNN selection and reconstruction pipeline.
For an air-shower search, however, only a small fraction of such events would accidentally align with the IceTop direction.

The expected number of false positives in the air-shower search was estimated using the formula:
\begin{equation}
    \text{FP}_1 = \frac{N_{\rm Pass}}{N_B} \times N_S~,
\end{equation}
where $N_B$ is the total number of FRT events, $N_{\rm Pass}$ is the number of FRT events passing the SNR or CNN quality cuts and successfully reconstructed, and  $N_S$  represents the total number of scintillator-triggered events. 
This yields a first-order estimate of false positives before imposing a directional consistency requirement with IceTop.

To obtain the final false positive rate after applying the $\Delta\omega$ cut, we calculate for each of the $N_{\rm Pass}$ FRT events the fraction $f_i^{\rm IT}$ of IceTop events within $7^\circ$ of the given FRT event. 
The total number of expected false positives is then:

\begin{equation}
   \text{FP}_{\rm tot} = \text{mean}(f_i^{\rm IT}) \times \text{FP}_1~.
\end{equation}

Using this approach, the SNR method yields an expected 3.7 false positives, while the CNN method results in only 0.08 expected false positives. A summary of the FPR calculation is presented in Table~\ref{tab:FPcomparison}.

This demonstrates that the use of CNNs not only lowers the detection threshold but at the same time significantly decreases the identification of false positive events.
Given the total livetime of the analysis, future analyses could adopt less conservative selection thresholds while still maintaining acceptable background rates, further highlighting the advantages of CNN-based selection over traditional methods.

\section{Conclusion}
\label{sec:outlook}

In this paper, we presented the development and successful implementation of deep-learning-based convolutional neural networks (CNNs), namely a \classifier\ and a \denoiser, applied to radio measurements of air showers at the IceCube Neutrino Observatory.
Both networks were trained on simulated CoREAS radio pulses added to measured background waveforms from a prototype station of IceCube’s surface array enhancement.

To experimentally test the performance of the CNNs, the networks were applied to air-shower events recorded over 106 days, triggered by the scintillation panels of the prototype station. The results were compared to a reference SNR method.

For both methods, the reconstructed arrival directions and comparisons with IceTop energy estimators confirm that we detect the radio pulses originating from air showers. 
This enables a direct comparison between the CNN and SNR methods, not only with simulated data, but also with these real air-shower measurements.

The CNN-based approach offers multiple advantages, particularly at low and intermediate SNR:
\begin{itemize}
    \item The CNNs improve the precision and bias of the reconstructed radio pulse power and arrival time, as confirmed on an independent validation dataset. At high SNR values, where the uncertainties are already low, the CNNs perform comparably to the reference SNR method, which makes them most useful for waveforms with low and intermediate SNRs.
    \item The CNNs increased the number of detected air-shower events by nearly a factor of five compared to the SNR method. Comparison with IceTop shows that most of the additionally identified events are of lower energy than those identified with the SNR method.
    \item Despite applying the same 95\% background rejection per antenna, the CNNs yielded substantially fewer false positives than the SNR method. This indicates that false positives in the SNR method are more likely due to correlated background across antennas, whereas the CNNs are more robust against such correlations.
\end{itemize}

Several insights emerged during network development that are relevant for future applications in other radio arrays.
For training the networks, it was important to use the background recorded by the same antennas and data acquisition system to which the networks would subsequently be applied. 
Training on the simulated background proved insufficient, likely due to its oversimplified representation of realistic, complex background noise.
Moreover, the real background also varied between antennas, e.g., because anthropogenic RFI depends on the orientation and position of the antennas and because the electronics response of the data-acquisition varies slightly from channel to channel.
Additionally, leveraging both polarization channels simultaneously improved network performance, as this captures the real correlations in the background of both channels. Multi-antenna correlations of both the background and the signal of an event can be considered in future work. This would need a drastic increase in computational resources, as noise and signal show different correlations, adding another dimension to the phase space. Further improvement of the \denoiser\ might be possible, e.g., by upsampling of the input traces~\cite{galvez_molina_2025_14919724}, which can be achieved with a modest increase of computational resources.

Finally, the selection criteria based on network outputs, can be adjusted for different needs. For example, if a particular physics analysis demands a purer data sample, a higher classifier score can be set to further reduce the false positive rate. This flexibility could be particularly useful for self-triggering radio experiments like GRAND~\cite{GRAND:2018iaj} or PUEO~\cite{PUEO:2020bnn}, which require low FPR.
Radio setups that are triggered by other air-shower detectors, like the IceTop enhancement, can typically tolerate a higher FPR and will primarily benefit from the lower detection threshold achieved with the CNNs. 
As demonstrated with the three prototype antennas of IceCube's surface enhancement, the CNNs enable detection of air showers down to a few tens of PeV.
We thus expect that they will play an important role for radio detection with the IceCube-Gen2 surface array~\cite{Schroder:2023lux}, as well as for other externally triggered radio experiments, such as at the Pierre Auger Observatory~\cite{PierreAuger:2018pmw,PierreAuger:2023gql,Verpoest:2024gsd} or the POEMMA Balloon-Radio mission~\cite{Battisti:2024jjy}. 

In conclusion, the neural networks developed in this work are a valuable tool for improving radio-based cosmic-ray detection. These networks can enhance the science reach of existing radio arrays, e.g., by lowering their detection threshold, and will contribute to achieving the science goals envisioned for future experiments.

\begin{acknowledgments}
The IceCube collaboration acknowledges the significant contributions to this manuscript from Alan Coleman, Abdul Rehman, and Frank Schroeder. We thank the Tunka-Rex Collaboration, in particular, Dmitriy Kostunin and Dmitry Shipilov, for making their code available, which became the initial basis for this study.
The authors gratefully acknowledge the support from the following agencies and institutions:
USA {\textendash} U.S. National Science Foundation-Office of Polar Programs,
U.S. National Science Foundation-Physics Division,
U.S. National Science Foundation-EPSCoR,
U.S. National Science Foundation-Office of Advanced Cyberinfrastructure,
National Aeronautics and Space Administration-EPSCoR,
Wisconsin Alumni Research Foundation,
University of Delaware high-performance computing resources,
Center for High Throughput Computing (CHTC) at the University of Wisconsin{\textendash}Madison,
Open Science Grid (OSG),
Partnership to Advance Throughput Computing (PATh),
Advanced Cyberinfrastructure Coordination Ecosystem: Services {\&} Support (ACCESS),
Frontera and Ranch computing project at the Texas Advanced Computing Center,
U.S. Department of Energy-National Energy Research Scientific Computing Center,
Particle astrophysics research computing center at the University of Maryland,
Institute for Cyber-Enabled Research at Michigan State University,
Astroparticle physics computational facility at Marquette University,
NVIDIA Corporation,
and Google Cloud Platform;
Belgium {\textendash} Funds for Scientific Research (FRS-FNRS and FWO),
FWO Odysseus and Big Science programmes,
and Belgian Federal Science Policy Office (Belspo);
Germany {\textendash} Bundesministerium f{\"u}r Bildung und Forschung (BMBF),
Deutsche Forschungsgemeinschaft (DFG),
European Research Council (ERC),
Helmholtz Alliance for Astroparticle Physics (HAP),
Initiative and Networking Fund of the Helmholtz Association,
Deutsches Elektronen Synchrotron (DESY),
and High Performance Computing cluster of the RWTH Aachen;
Sweden {\textendash} Swedish Research Council,
Swedish Polar Research Secretariat,
Swedish National Infrastructure for Computing (SNIC),
and Knut and Alice Wallenberg Foundation;
European Union {\textendash} EGI Advanced Computing for research;
Australia {\textendash} Australian Research Council;
Canada {\textendash} Natural Sciences and Engineering Research Council of Canada,
Calcul Qu{\'e}bec, Compute Ontario, Canada Foundation for Innovation, WestGrid, and Digital Research Alliance of Canada;
Denmark {\textendash} Villum Fonden, Carlsberg Foundation, and European Commission;
New Zealand {\textendash} Marsden Fund;
Japan {\textendash} Japan Society for Promotion of Science (JSPS)
and Institute for Global Prominent Research (IGPR) of Chiba University;
Korea {\textendash} National Research Foundation of Korea (NRF);
Switzerland {\textendash} Swiss National Science Foundation (SNSF).
\end{acknowledgments}

\bibliography{biblio}

@PREAMBLE{
 "\providecommand{\noopsort}[1]{}" 
 # "\providecommand{\singleletter}[1]{#1}%" 
}

@article{galvez_molina_2025_14919724,
  author       = {Gálvez Molina, Paula and others},
  collaboration = "IceCube",
  title        = "{Detection of Radio Signals from Cosmic Rays Using Convolutional Neural Networks with Data from SKALA antennas at IceTop}",
  journal = "{Workshop on Machine Learning for Analysis of High-Energy Cosmic Particles, 27-31 January 2025, University of Delaware}",
  year         = 2025,
  publisher    = {Zenodo},
  doi          = {10.5281/zenodo.14919724},
  url          = {https://doi.org/10.5281/zenodo.14919724},
}

@article{GRAND:2018iaj,
    author = "\'Alvarez-Mu\~niz, Jaime and others",
    collaboration = "GRAND",
    title = "{The Giant Radio Array for Neutrino Detection (GRAND): Science and Design}",
    eprint = "1810.09994",
    archivePrefix = "arXiv",
    primaryClass = "astro-ph.HE",
    doi = "10.1007/s11433-018-9385-7",
    journal = "Sci. China Phys. Mech. Astron.",
    volume = "63",
    number = "1",
    pages = "219501",
    year = "2020"
}

@article{PUEO:2020bnn,
    author = "Abarr, Q. and others",
    collaboration = "PUEO",
    title = "{The Payload for Ultrahigh Energy Observations (PUEO): a white paper}",
    eprint = "2010.02892",
    archivePrefix = "arXiv",
    primaryClass = "astro-ph.IM",
    doi = "10.1088/1748-0221/16/08/P08035",
    journal = "JINST",
    volume = "16",
    number = "08",
    pages = "P08035",
    year = "2021"
}

@article{Battisti:2024jjy,
    author = "Battisti, Matteo and Eser, Johannes and Olinto, Angela and Osteria, Giuseppe",
    collaboration = "JEM-EUSO",
    title = "{POEMMA-Balloon with Radio: A balloon-born multi-messenger multi-detector observatory}",
    eprint = "2409.06753",
    archivePrefix = "arXiv",
    primaryClass = "astro-ph.IM",
    doi = "10.1016/j.nima.2024.169819",
    journal = "Nucl. Instrum. Meth. A",
    volume = "1069",
    pages = "169819",
    year = "2024"
}

@article{Schroder:2023lux,
    author = {Schr\"oder, Frank G.},
    collaboration = "IceCube-Gen2",
    title = "{Design and Expected Performance of the IceCube-Gen2 Surface Array and its Radio Component}",
    eprint = "2306.05900",
    archivePrefix = "arXiv",
    primaryClass = "astro-ph.HE",
    doi = "10.22323/1.424.0058",
    journal = "PoS",
    volume = "ARENA2022",
    pages = "058",
    year = "2023"
}

@article{Verpoest:2024gsd,
 author = "{S. Verpoest \textit{et al.}}",
 collaboration = "IceCube-Gen2 and Pierre Auger",

 title = "{Observation of air showers with an IceCube-Gen2 prototype station at the Pierre Auger Observatory}",
 doi = "10.22323/1.470.0037",
 eprint = "2409.00713",
 archivePrefix = "arXiv",
 primaryClass = "astro-ph.HE",
 journal = "PoS",
 year = 2024,
 volume = "ARENA2024",
 pages = "037"
}

@article{PierreAuger:2023gql,
    author = "Abdul Halim, Adila and others",
    collaboration = "Pierre Auger",
    title = "{Status and expected performance of the AugerPrime Radio Detector}",
    doi = "10.22323/1.444.0344",
    journal = "PoS",
    volume = "ICRC2023",
    pages = "344",
    year = "2023"
}

@article{PierreAuger:2018pmw,
    author = "Aab, Alexander and others",
    collaboration = "Pierre Auger",
    title = "{Observation of inclined EeV air showers with the radio detector of the Pierre Auger Observatory}",
    eprint = "1806.05386",
    archivePrefix = "arXiv",
    primaryClass = "astro-ph.IM",
    reportNumber = "FERMILAB-PUB-18-259-ND",
    doi = "10.1088/1475-7516/2018/10/026",
    journal = "JCAP",
    volume = "10",
    pages = "026",
    year = "2018"
}

@misc{franccois2019keras,
  title={Keras},
  author={Chollet, Fran\c{c}ois and others},
  year={2015},
  howpublished={\url{https://keras.io}},
}

@article{abadi2016tensorflow,
  title={Tensorflow: Large-scale machine learning on heterogeneous distributed systems},
  author={Abadi, Mart{\'\i}n and others},
  journal={arXiv preprint arXiv:1603.04467},
  year={2016}
}

@ARTICLE{2014arXiv1412.6071G,
       author = {{Graham}, Benjamin},
        title = "{Fractional Max-Pooling}",
      journal = {arXiv e-prints},
         year = 2014,
        month = dec,
          eid = {arXiv:1412.6071},
        pages = {arXiv:1412.6071},
          doi = {10.48550/arXiv.1412.6071},
archivePrefix = {arXiv},
       eprint = {1412.6071},
 primaryClass = {cs.CV},
       adsurl = {https://ui.adsabs.harvard.edu/abs/2014arXiv1412.6071G}
}

@article{IceCube:2012nn,
    author = "Abbasi, R. and others",
    collaboration = "IceCube",
    title = "{IceTop: The surface component of IceCube}",
    eprint = "1207.6326",
    archivePrefix = "arXiv",
    primaryClass = "astro-ph.IM",
    doi = "10.1016/j.nima.2012.10.067",
    journal = "Nucl. Instrum. Meth. A",
    volume = "700",
    pages = "188--220",
    year = "2013"
}

@article{Erdmann:2019nie,
    author = {Erdmann, M. and Schl\"uter, F. and Smida, R.},
    title = "{Classification and Recovery of Radio Signals from Cosmic Ray Induced Air Showers with Deep Learning}",
    eprint = "1901.04079",
    archivePrefix = "arXiv",
    primaryClass = "astro-ph.IM",
    doi = "10.1088/1748-0221/14/04/P04005",
    journal = "JINST",
    volume = "14",
    number = "04",
    pages = "P04005",
    year = "2019"
}

@article{LOPES:2015eya,
    author = "Apel, W. D. and others",
    collaboration = "LOPES",
    title = "{Improved absolute calibration of LOPES measurements and its impact on the comparison with REAS 3.11 and CoREAS simulations}",
    eprint = "1507.07389",
    archivePrefix = "arXiv",
    primaryClass = "astro-ph.HE",
    doi = "10.1016/j.astropartphys.2015.09.002",
    journal = "Astropart. Phys.",
    volume = "75",
    pages = "72--74",
    year = "2016"
}

@article{T-510:2015pyu,
    author = "Belov, K. and others",
    collaboration = "T-510",
    title = "{Accelerator measurements of magnetically-induced radio emission from particle cascades with applications to cosmic-ray air showers}",
    eprint = "1507.07296",
    archivePrefix = "arXiv",
    primaryClass = "astro-ph.IM",
    reportNumber = "SLAC-PUB-16338",
    doi = "10.1103/PhysRevLett.116.141103",
    journal = "Phys. Rev. Lett.",
    volume = "116",
    number = "14",
    pages = "141103",
    year = "2016"
}

@article{Tunka-Rex:2016nto,
    author = "Apel, W. D. and others",
    collaboration = "Tunka-Rex, LOPES",
    title = "{A comparison of the cosmic-ray energy scales of Tunka-133 and KASCADE-Grande via their radio extensions Tunka-Rex and LOPES}",
    eprint = "1610.08343",
    archivePrefix = "arXiv",
    primaryClass = "astro-ph.IM",
    doi = "10.1016/j.physletb.2016.10.031",
    journal = "Phys. Lett. B",
    volume = "763",
    pages = "179--185",
    year = "2016"
}

@misc{hepmllivingreview,
    Author = "{HEP ML Community}",
    title = "{A Living Review of Machine Learning for Particle Physics}",
    url={https://iml-wg.github.io/HEPML-LivingReview/},
    note={\url{https://iml-wg.github.io/HEPML-LivingReview/}, accessed on 28 Nov 2025}
}

@article{Bezyazeekov:2015rpa,
    author = "Bezyazeekov, P. A. and others",
    collaboration = "Tunka-Rex",
    title = "{Measurement of cosmic-ray air showers with the Tunka Radio Extension (Tunka-Rex)}",
    eprint = "1509.08624",
    archivePrefix = "arXiv",
    primaryClass = "astro-ph.IM",
    doi = "10.1016/j.nima.2015.08.061",
    journal = "Nucl. Instrum. Meth. A",
    volume = "802",
    pages = "89--96",
    year = "2015"
}

@article{Bezyazeekov:2021sha,
    author = "Bezyazeekov, Pavel and others",
    collaboration = "Tunka-Rex",
    title = "{Reconstruction of sub-threshold events of cosmic-ray radio detectors using an autoencoder}",
    eprint = "2108.04627",
    archivePrefix = "arXiv",
    primaryClass = "astro-ph.IM",
    doi = "10.22323/1.395.0223",
    journal = "PoS",
    volume = "ICRC2021",
    pages = "223",
    year = "2021"
}

@article{Shipilov:2018wph,
    author = "Shipilov, D. and others",
    collaboration = "Tunka-Rex",
    editor = "Riccobene, G. and Biagi, S. and Capone, A. and Distefano, C. and Piattelli, P.",
    title = "{Signal recognition and background suppression by matched filters and neural networks for Tunka-Rex}",
    eprint = "1812.03347",
    archivePrefix = "arXiv",
    primaryClass = "astro-ph.IM",
    doi = "10.1051/epjconf/201921602003",
    journal = "EPJ Web Conf.",
    volume = "216",
    pages = "02003",
    year = "2019"
}

@article{IceCube:2021epf,
    author = "Abbasi, Rasha and others",
    collaboration = "IceCube",
    title = "{First air-shower measurements with the prototype station of the IceCube surface enhancement}",
    eprint = "2107.08750",
    archivePrefix = "arXiv",
    primaryClass = "astro-ph.HE",
    reportNumber = "PoS-ICRC2021-314",
    doi = "10.22323/1.395.0314",
    journal = "PoS",
    volume = "ICRC2021",
    pages = "314",
    year = "2021"
}

@article{Paudel:2024wtm,
    author = "Paudel, Ek Narayan",
    collaboration = "IceCube",
    title = "{New IceTop trigger in the context of the planned IceCube surface detector enhancement at the South Pole}",
    eprint = "2401.12026",
    archivePrefix = "arXiv",
    primaryClass = "astro-ph.HE",
    doi = "10.1088/1748-0221/19/01/C01031",
    journal = "JINST",
    volume = "19",
    number = "01",
    pages = "C01031",
    year = "2024"
}

@article{benthem2021aperture,
  title={The aperture array verification system 1: system overview and early commissioning results},
  author={Benthem, P and Wayth, R and de Lera Acedo, E and Adami, K Zarb and Alderighi, MONICA and Belli, Carolina and Bolli, Pietro and Booler, T and Borg, Josef and Broderick, JW and others},
  journal={Astronomy \& Astrophysics},
  volume={655},
  pages={A5},
  year={2021},
  publisher={EDP Sciences}
}

@phdthesis{Turcotte-Tardif:2022smf,
    author = "Turcotte-Tardif, Roxanne",
    title = "{Radio measurements of cosmic rays at the South Pole}",
    doi = "10.5445/IR/1000160782",
    school = "KIT, Karlsruhe",
    year = "2022",
    note = {\url{https://publikationen.bibliothek.kit.edu/1000160782}},
}

@phdthesis{A-Rehman:2024,
    author = "Rehman, Abdul",
    title = "{Convolutional Neural Networks for Radio Signals from Cosmic-Ray Air Showers}",
    school = "University of Delaware, ProQuest Dissertations",
    year = "2024",
    note = {\url{https://www.proquest.com/dissertations-theses/convolutional-neural-networks-radio-signals/docview/3163000979/se-2}},
}

@article{IceCube:2023pjc,
    author = "Schr{\"o}der, Frank G. and others",
    collaboration = "IceCube",
    title = "{Status and plans for the instrumentation of the IceCube Surface Array Enhancement}",
    eprint = "2402.17854",
    archivePrefix = "arXiv",
    primaryClass = "astro-ph.IM",
    reportNumber = "PoS-ICRC2023-342",
    doi = "10.22323/1.444.0342",
    journal = "PoS",
    volume = "ICRC2023",
    pages = "342",
    year = "2023"
}

@article{Schroder:2016hrv,
    author = {Schr\"oder, Frank G.},
    title = "{Radio detection of Cosmic-Ray Air Showers and High-Energy Neutrinos}",
    eprint = "1607.08781",
    archivePrefix = "arXiv",
    primaryClass = "astro-ph.IM",
    doi = "10.1016/j.ppnp.2016.12.002",
    journal = "Prog. Part. Nucl. Phys.",
    volume = "93",
    pages = "1--68",
    year = "2017"
}

@article{Huege:2016veh,
    author = "Huege, T.",
    title = "{Radio detection of cosmic ray air showers in the digital era}",
    eprint = "1601.07426",
    archivePrefix = "arXiv",
    primaryClass = "astro-ph.IM",
    doi = "10.1016/j.physrep.2016.02.001",
    journal = "Phys. Rept.",
    volume = "620",
    pages = "1--52",
    year = "2016"
}

@article{LOPES:2021ipp,
    author = "Apel, W. D. and others",
    collaboration = "LOPES",
    title = "{Final results of the LOPES radio interferometer for cosmic-ray air showers}",
    eprint = "2102.03928",
    archivePrefix = "arXiv",
    primaryClass = "astro-ph.HE",
    doi = "10.1140/epjc/s10052-021-08912-4",
    journal = "Eur. Phys. J. C",
    volume = "81",
    number = "2",
    pages = "176",
    year = "2021"
}

@article{Buitink:2014eqa,
    author = "Buitink, S. and others",
    collaboration = {LOFAR},
    title = "{Method for high precision reconstruction of air shower $X_{max}$ using two-dimensional radio intensity profiles}",
    eprint = "1408.7001",
    archivePrefix = "arXiv",
    primaryClass = "astro-ph.IM",
    doi = "10.1103/PhysRevD.90.082003",
    journal = "Phys. Rev. D",
    volume = "90",
    number = "8",
    pages = "082003",
    year = "2014"
}

@article{PierreAuger:2016vya,
    author = "Aab, Alexander and others",
    collaboration = "Pierre Auger",
    title = "{Measurement of the Radiation Energy in the Radio Signal of Extensive Air Showers as a Universal Estimator of Cosmic-Ray Energy}",
    eprint = "1605.02564",
    archivePrefix = "arXiv",
    primaryClass = "astro-ph.HE",
    reportNumber = "FERMILAB-PUB-16-169-AD-AE-CD-TD",
    doi = "10.1103/PhysRevLett.116.241101",
    journal = "Phys. Rev. Lett.",
    volume = "116",
    number = "24",
    pages = "241101",
    year = "2016"
}

@article{Bezyazeekov:2018yjw,
    author = "Bezyazeekov, P. A. and others",
    collaboration = {Tunka-Rex},
    title = "{Reconstruction of cosmic ray air showers with Tunka-Rex data using template fitting of radio pulses}",
    eprint = "1803.06862",
    archivePrefix = "arXiv",
    primaryClass = "astro-ph.IM",
    doi = "10.1103/PhysRevD.97.122004",
    journal = "Phys. Rev. D",
    volume = "97",
    number = "12",
    pages = "122004",
    year = "2018"
}

@article{BalagopalV:2017aan,
    author = "Balagopal V., A. and Haungs, A. and Huege, T. and Schr{\"o}der, F. G.",
    title = "{Search for PeVatrons at the Galactic Center using a radio air-shower array at the South Pole}",
    eprint = "1712.09042",
    archivePrefix = "arXiv",
    primaryClass = "astro-ph.IM",
    doi = "10.1140/epjc/s10052-018-5537-2",
    journal = "Eur. Phys. J. C",
    volume = "78",
    number = "2",
    pages = "111",
    year = "2018",
    note = "[Erratum: Eur.Phys.J.C 78, 1017 (2018), Erratum: Eur.Phys.J.C 81, 483 (2021)]"
}

@article{ANITA:2010ect,
    author = "Hoover, S. and others",
    collaboration = "ANITA",
    title = "{Observation of Ultra-high-energy Cosmic Rays with the ANITA Balloon-borne Radio Interferometer}",
    eprint = "1005.0035",
    archivePrefix = "arXiv",
    primaryClass = "astro-ph.HE",
    doi = "10.1103/PhysRevLett.105.151101",
    journal = "Phys. Rev. Lett.",
    volume = "105",
    pages = "151101",
    year = "2010"
}

@article{Schoorlemmer:2020low,
    author = "Schoorlemmer, Harm and Carvalho, Washington R.",
    title = "{Radio interferometry applied to the observation of cosmic-ray induced extensive air showers}",
    eprint = "2006.10348",
    archivePrefix = "arXiv",
    primaryClass = "astro-ph.HE",
    doi = "10.1140/epjc/s10052-021-09925-9",
    journal = "Eur. Phys. J. C",
    volume = "81",
    number = "12",
    pages = "1120",
    year = "2021"
}

@article{PierreAuger:2023lkx,
    author = "Abdul Halim, A. and others",
    collaboration = "Pierre Auger",
    title = "{Demonstrating Agreement between Radio and Fluorescence Measurements of the Depth of Maximum of Extensive Air Showers at the Pierre Auger Observatory}",
    eprint = "2310.19963",
    archivePrefix = "arXiv",
    primaryClass = "astro-ph.HE",
    reportNumber = "FERMILAB-PUB-24-0138-AD-CSAID-PPD-TD-V",
    doi = "10.1103/PhysRevLett.132.021001",
    journal = "Phys. Rev. Lett.",
    volume = "132",
    number = "2",
    pages = "021001",
    year = "2024"
}

@article{Holt:2019fnj,
    author = {Holt, Ewa M. and Schr\"oder, Frank G. and Haungs, Andreas},
    title = "{Enhancing the cosmic-ray mass sensitivity of air-shower arrays by combining radio and muon detectors}",
    eprint = "1905.01409",
    archivePrefix = "arXiv",
    primaryClass = "astro-ph.HE",
    doi = "10.1140/epjc/s10052-019-6859-4",
    journal = "Eur. Phys. J. C",
    volume = "79",
    number = "5",
    pages = "371",
    year = "2019"
}

@article{Flaggs:2023exc,
    author = {Flaggs, Benjamin and Coleman, Alan and Schr\"oder, Frank G.},
    title = "{Studying the mass sensitivity of air-shower observables using simulated cosmic rays}",
    eprint = "2306.13246",
    archivePrefix = "arXiv",
    primaryClass = "hep-ph",
    doi = "10.1103/PhysRevD.109.042002",
    journal = "Phys. Rev. D",
    volume = "109",
    number = "4",
    pages = "042002",
    year = "2024"
}

@article{Andringa:2011zz,
    author = "Andringa, S. and Conceicao, R. and Pimenta, M.",
    title = "{Mass composition and cross-section from the shape of cosmic ray shower longitudinal profiles}",
    doi = "10.1016/j.astropartphys.2010.10.002",
    journal = "Astropart. Phys.",
    volume = "34",
    pages = "360--367",
    year = "2011"
}

@article{KASCADE:2005ynk,
    author = "Antoni, T. and others",
    collaboration = "KASCADE",
    title = "{KASCADE measurements of energy spectra for elemental groups of cosmic rays: Results and open problems}",
    eprint = "astro-ph/0505413",
    archivePrefix = "arXiv",
    doi = "10.1016/j.astropartphys.2005.04.001",
    journal = "Astropart. Phys.",
    volume = "24",
    pages = "1--25",
    year = "2005"
}

@article{Coleman:2022abf,
    author = "Coleman, A. and others",
    title = "{Ultra high energy cosmic rays The intersection of the Cosmic and Energy Frontiers}",
    eprint = "2205.05845",
    archivePrefix = "arXiv",
    primaryClass = "astro-ph.HE",
    reportNumber = "FERMILAB-PUB-22-413-PPD",
    doi = "10.1016/j.astropartphys.2023.102819",
    journal = "Astropart. Phys.",
    volume = "149",
    pages = "102819",
    year = "2023"
}

@article{heck1998corsika,
  title={{CORSIKA}: A Monte Carlo code to simulate extensive air showers},
  author={Heck, Dieter and others},
  journal={{FZKA} Report},
  volume={6019},
  number={11},
  year={1998},
  publisher={Citeseer}
}

@article{huege2013simulating,
    author = "Huege, T. and Ludwig, M. and James, C. W.",
    editor = "Lahmann, Robert and Eberl, Thomas and Graf, Kay and James, Clancy and Huege, Tim and Karg, Timo and Nahnhauer, Rolf",
    title = "{Simulating radio emission from air showers with CoREAS}",
    eprint = "1301.2132",
    archivePrefix = "arXiv",
    primaryClass = "astro-ph.HE",
    doi = "10.1063/1.4807534",
    journal = "AIP Conf. Proc.",
    volume = "1535",
    number = "1",
    pages = "128",
    year = "2013"
}

@article{BOHLEN2014211,
    title = {The FLUKA Code: Developments and Challenges for High Energy and Medical Applications},
    journal = {Nuclear Data Sheets},
    volume = {120},
    pages = {211-214},
    year = {2014},
    issn = {0090-3752},
    doi = {https://doi.org/10.1016/j.nds.2014.07.049},
    url = {https://www.sciencedirect.com/science/article/pii/S0090375214005018},
    author = {T.T. Böhlen and F. Cerutti and M.P.W. Chin and A. Fassò and A. Ferrari and P.G. Ortega and A. Mairani and P.R. Sala and G. Smirnov and V. Vlachoudis}
}

@article{Riehn:2019jet,
    author = "Riehn, Felix and Engel, Ralph and Fedynitch, Anatoli and Gaisser, Thomas K. and Stanev, Todor",
    title = "{Hadronic interaction model Sibyll 2.3d and extensive air showers}",
    eprint = "1912.03300",
    archivePrefix = "arXiv",
    primaryClass = "hep-ph",
    doi = "10.1103/PhysRevD.102.063002",
    journal = "Phys. Rev. D",
    volume = "102",
    number = "6",
    pages = "063002",
    year = "2020"
}

@article{IceCube:2023nyj,
    author = "Abbasi, Rasha and others",
    collaboration = "IceCube",
    title = "{Accounting for changing snow over 10 years of IceTop, and its impact on the all-particle cosmic ray spectrum}",
    doi = "10.22323/1.444.0377",
    journal = "PoS",
    volume = "ICRC2023",
    pages = "377",
    year = "2023"
}

@article{Leszczynska:2019ahq,
    author = "Leszczy\'nska, Agnieszka and Plum, Matthias",
    collaboration = "IceCube",
    title = "{Simulation and Reconstruction Study of a Future Surface Scintillator Array at the IceCube Neutrino Observatory}",
    eprint = "1909.02258",
    archivePrefix = "arXiv",
    primaryClass = "astro-ph.IM",
    reportNumber = "PoS-ICRC2019-332",
    doi = "10.22323/1.358.0332",
    journal = "PoS",
    volume = "ICRC2019",
    pages = "332",
    year = "2020"
}

@article{Haungs:2019ylq,
    author = "Haungs, Andreas",
    editor = "Lhenry-Yvon, I. and Biteau, J. and Biteau, O. and Ghia, P.",
    collaboration = "IceCube",
    title = "{A Scintillator and Radio Enhancement of the IceCube Surface Detector Array}",
    eprint = "1903.04117",
    archivePrefix = "arXiv",
    primaryClass = "astro-ph.IM",
    doi = "10.1051/epjconf/201921006009",
    journal = "EPJ Web Conf.",
    volume = "210",
    pages = "06009",
    year = "2019"
}

@article{Schroder:2018dvb,
    author = {Schr\"oder, Frank G.},
    editor = "Riccobene, G. and Biagi, S. and Capone, A. and Distefano, C. and Piattelli, P.",
    collaboration = "IceCube-Gen2",
    title = "{Physics Potential of a Radio Surface Array at the South Pole}",
    eprint = "1811.00599",
    archivePrefix = "arXiv",
    primaryClass = "astro-ph.IM",
    doi = "10.1051/epjconf/201921601007",
    journal = "EPJ Web Conf.",
    volume = "216",
    pages = "01007",
    year = "2019"
}

@INPROCEEDINGS{7297231,
  author={de Lera Acedo, E. and others},
  booktitle={2015 International Conference on Electromagnetics in Advanced Applications (ICEAA)}, 
  title={Evolution of {SKALA} ({SKALA-2}), the log-periodic array antenna for the SKA-low instrument}, 
  year={2015},
  volume={},
  number={},
  pages={839-843},
  doi={10.1109/ICEAA.2015.7297231}
}

@article{IceCube:2022dcd,
    author = "Abbasi, R. and others",
    collaboration = "IceCube",
    title = "{Framework and tools for the simulation and analysis of the radio emission from air showers at IceCube}",
    eprint = "2205.02258",
    archivePrefix = "arXiv",
    primaryClass = "astro-ph.HE",
    doi = "10.1088/1748-0221/17/06/P06026",
    journal = "JINST",
    volume = "17",
    number = "06",
    pages = "P06026",
    year = "2022"
}

@article{Rehman:2023jme,
    author = {Rehman, Abdul and Coleman, Alan and Schr\"oder, Frank G.},
    collaboration = "IceCube",
    title = "{Deep Learning for the classification and recovery of Cosmic-Ray signals against background measured at South Pole}",
    doi = "10.22323/1.424.0012",
    journal = "PoS",
    volume = "ARENA2022",
    pages = "012",
    year = "2023"
}

@article{Abbasi:2024ws,
    author = {Abbasi, R. and others},
    collaboration = "IceCube",
    title = "{Enhancing air-shower observations: Results from an IceCube Surface Array Prototype Station}",
    doi = "10.22323/1.470.0038",
    journal = "PoS",
    volume = "ARENA2024",
    pages = "038",
    year = "2024"
}

@article{Schroder:2023sam,
    author = {Schr\"oder, Frank G. and Connolly, Amy L. and Huege, Tim and Rehman, Abdul},
    title = "{Towards a Standard Definition of the Signal-to-Noise Ratio for Radio Signals of ultra-high-energy Particles}",
    eprint = "2306.05901",
    archivePrefix = "arXiv",
    primaryClass = "astro-ph.IM",
    doi = "10.22323/1.424.0027",
    journal = "PoS",
    volume = "ARENA2022",
    pages = "027",
    year = "2023"
}

\end{document}